\documentclass[fleqn,usenatbib]{mnras}
\usepackage{newtxtext,newtxmath}

\usepackage[T1]{fontenc}

\DeclareRobustCommand{\VAN}[3]{#2}
\let\VANthebibliography\thebibliography
\def\thebibliography{\DeclareRobustCommand{\VAN}[3]{##3}\VANthebibliography}

\usepackage{graphicx}	
\usepackage{amsmath}	
\usepackage{threeparttable}
\usepackage{xcolor}

\graphicspath{{figures/}}

\title[]{Tracing Milky Way scattering by compact extragalactic radio sources}

\author[Koryukova et al.]{\parbox{\textwidth}{
T. A. Koryukova,$^{1}$\thanks{E-mail: tatyana.koryukova@gmail.com}
A. B. Pushkarev,$^{2,1}$
A. V. Plavin,$^{1,3}$
Y. Y. Kovalev$^{1,3,4}$
}
\vspace{0.4cm}\\
\parbox{\textwidth}{
$^1$Lebedev Physical Institute of the Russian Academy of Sciences, Leninsky prospekt 53, 119991 Moscow, Russia\\
$^2$Crimean Astrophysical Observatory, Nauchny 298409, Crimea, Russia\\
$^3$Moscow Institute of Physics and Technology, Institutsky per. 9, Dolgoprudny 141700, Russia\\
$^4$Max-Planck-Institut f\"ur Radioastronomie, Auf dem H\"ugel 69, 53121 Bonn, Germany
}}

\date{Accepted 2022 July 01. Received 2022 June 29; in original form 2021 December 30}

\pubyear{2022}

\begin{document}
\label{firstpage}
\pagerange{\pageref{firstpage}--\pageref{lastpage}}
\maketitle

\begin{abstract}
We used archival very long baseline interferometry (VLBI) data of active galactic nuclei (AGN) observed from 1.4~GHz to 86~GHz to measure the angular size of VLBI radio cores in 8959 AGNs. We analysed their sky distributions, frequency dependencies and created the most densely sampled and complete to date distribution map of large-scale scattering properties of the interstellar medium in our Galaxy. Signiﬁcant angular broadening of the measured AGN core sizes is detected for the sources seen through the Galactic plane, and this eﬀect is especially strong at low frequencies (e.g.,\ at 2~GHz). The scattering screens containing electron density ﬂuctuations of hot plasma are mainly concentrated in the Galactic plane and manifest clumpy distribution. The region of the strongest scattering is the Galactic centre, where the Galactic bar and the compact radio source Sagittarius~A$^\ast$ are located. We have also found the enhancement of scattering strength in regions of the Cygnus constellation, supernova remnants Taurus A, Vela, W78 and Cassiopeia~A, and the Orion Nebula. Using multi-frequency observational data of AGN core sizes, we separated the contribution of the intrinsic and scattered sizes to the measured angular diameter for 1411 sources. For the sources observed through the Galactic plane, the contribution of the scattered size component is systematically larger than for those seen outside the Galactic plane. The derived power-law scattering indices are found to be in good agreement with theoretical predictions for the diffractive-dominated scattering of radio emission in a hot plasma with Gaussian distribution of density inhomogeneities.

\end{abstract}



\section{Introduction}
\label{ch:intro}

The interstellar medium is a rarefied medium which fills the space between stars in galaxies. The ISM includes interstellar gas (molecular, atomic, ionized), dust, electromagnetic fields and cosmic rays. All the components of the ISM are closely related to each other due to the constant circulation of matter and energy in the galaxy. The ISM is highly turbulent \citep{TurbISM2014}, with high Reynolds numbers \citep[$R_e\gg10^3$, e.g.,][]{Combes2000}. Turbulence can be caused by processes on a wide range of scales (from pc to kpc), e.g., the interaction of cosmic rays and the interstellar plasma, stars and the interstellar medium, as well as rotation of the Galaxy, collisions between the stellar and gas components \citep{Elmegreen_2004}, etc.

When a radio wave passes through turbulent ionized gas of the ISM, it encounters stochastic free-electron density fluctuations on its way, which cause fluctuations in the refractive index of the medium. Thus, wavefronts will be randomly distorted, resulting in scattering of radio emission \citep{Ferrire2019}. Scattering may distort the images of compact radio sources in a number of ways \citep[e.g.,][]{2008A&A...489L..33S,Pushkarev2013,2014ApJ...794L..14G,2018ApJ...865..104J}, this complicates the interpretation of their observations. The study of radio wave scattering effects makes it possible to reconstruct the intrinsic characteristics of a scattered source. It is important to consider the phenomena of radio wave scattering when measuring the brightness temperatures of AGNs, which is crucial for theoretical models of relativistic jets \citep{Johnson2016}. At the same time, radiation scattering contains the most important information about properties of the turbulent medium, and a detailed study of this effect allows to investigate the properties of the ISM in our Galaxy \citep[e.g.,][]{Pushkarev15}. The AGN VLBI radio core (or shortly AGN core) that is generally observed in the region where the jet stops being opaque to synchrotron radiation is compact enough to probe the scattering properties of the ISM. The advantage of these sources over pulsars is that they are (i) more numerous, (ii) uniformly distributed over the sky, and (iii) their emission passes through the entire depth of the scattering screens in the Galaxy.

According to the studies of the radio wave scattering based on pulsar observations, it has been shown that scattering screens in our Galaxy consist of two main components: (1) a nearly uniform medium with a characteristic length-scale of the density fluctuations of about 500~pc and (2) a clumped medium with a scale of 100~pc, with an approximate size of one clump about 1~pc \citep{Cordes1985,Cordes86}. The typical size of such a substructure is important since it directly determines how the scattering manifests itself. For example, large-scale inhomogeneities lead to the refractive scattering effects, e.g., angular wandering of the apparent source position \citep{Clegg_1998}, extreme scattering effects \citep{Fiedler1987,Pushkarev2013}, `slow' intensity variations \citep{Rickett1984}. Fluctuations at smaller scales lead to diffractive scattering effects, e.g., angular broadening \citep{Duffett_Smith1976}, `fast' intensity scintillations in time and frequency \citep{Jauncey2020}.

There are two competing models of scattering screens: the Gaussian screen model \citep{Gauss_screen,Ratcliffe_1956} and the model with a power-law spectrum of electron density fluctuations \citep{Lovelace1970, Cordes1985}. The brightness distribution of a point source seen through the Gaussian scattering screen is Gaussian (unresolved core surrounded by a halo) with an angular size proportional to wavelength squared, i.e., $\theta\propto\lambda^{k}$, where $k = 2.0$ \citep{Narayan1985,Cordes86}. A power-law spectrum of turbulence assumes that the spectrum of free electron density fluctuations can be approximated as a Kolmogorov power-law \citep{Kolmogorov41,Cordes86,Rickett1990,Armstrong95}. In this case, the scattering angle of a point source is proportional to wavelength to the power of 2.2, i.e., $\theta\propto\lambda^{k}$, where $k = 2.2$.

If there is no intermediate scattering screen, then the observed angular size of a compact background source will coincide with its intrinsic size. For example, in the case of an apparent jet base in active galactic nuclei, an observed size will be proportional to $\lambda^{k}$, where $k = 1$ for conical jet shape \citep{BK79,Konigl1981}. This assumption works for 15~GHz and lower frequencies. At higher frequencies, VLBI observations of nearby AGNs might probe innermost jet regions characterized by quasi-parabolic shape \citep{Asada2012,Kovalev2020}.

In this work we further develop the study made by \cite{Pushkarev15}. For that, we use a larger sample of the observed AGNs, and introduce new methods for modelling the apparent jet base structure of active galactic nuclei and a completely new approach to study the scattering effects in our Galaxy.

\section{AGN VLBI core size measurements}
\label{ch:size_measurements}

Our analysis is based on the VLBI observations of AGN jets at frequencies ranging from 1.4 to 86~GHz (see \autoref{tab:table_freqs} for details) compiled in the Astrogeo database\footnote{\url{ http://astrogeo.org/vlbi_images/}}. We rely on the measured interferometric visibilities and do not analyse the corresponding restored images. The Astrogeo database collects geodetic VLBI observations \citep{2009JGeod..83..859P,2012A&A...544A..34P,2012ApJ...758...84P}, the VLBA\footnote{Very Long Baseline Array of the National Radio Astronomy Observatory, Socorro, NM, USA} calibrator surveys (VCS; \citealt{2002ApJS..141...13B,2003AJ....126.2562F,2005AJ....129.1163P,2006AJ....131.1872P,2007AJ....133.1236K,2008AJ....136..580P}), the MOJAVE VLBA program (\citealt{2018ApJS..234...12L} and references therein) and other VLBI networks including EVN\footnote{European VLBI Network}, LBA\footnote{Long Baseline Array} and GMVA\footnote{Global millimeter VLBI array} \citep{2007ApJ...658..203H,2008AJ....136..159L,2011AJ....142...35P,2011MNRAS.414.2528P,2011AJ....142..105P,2012MNRAS.419.1097P,2013AJ....146....5P,2015ApJS..217....4S,2017ApJS..230...13S,2017ApJ...846...98J,2019MNRAS.485...88P,2019A&A...622A..92N,2021AJ....161...14P,2021AJ....161...88P}. This dataset contains 17\,474 sources observed from 1994 to 2021, comprising more than $100\,000$ individual observations. The majority of them were performed at 2, 5, 8, or 15~GHz.

The AGN VLBI core is the apparent jet origin. We apply the model-fitting of interferometric visibilities approach and describe the apparent structure with two Gaussian components: the core and the extended jet emission. We use nested sampling to make fitting completely automatic and independent of initial guesses. The brightest component of the two is first selected as the core; if the jet orientation at a lower frequency is opposite to that at a higher frequency, we switch the two components. \cite{plavin2022} present and discuss this in details as well as evaluate this fitting approach used to measure jet directions. Nested sampling provides principled uncertainty estimates on all parameters, including the core component size. We find these uncertainties directly useful for our analysis even though they are fundamentally underestimated: calibration and self-calibration effects are not accounted here. Specifically, we drop the measurements with uncertainty in the core size exceeding 50~per~cent of the value itself. We find that the unresolved sources are also removed by this criterion and we do not perform any additional selection. \autoref{tab:exp_data} contains eight randomly selected AGN for which the core size was measured.

\begin{table}
	\caption{Central frequencies of the used observing bands.}
	\centering
	\begin{tabular}{|*{3}{c|}} 
		\hline
		\hline
		 Band   & Central frequency range (GHz) \\
		\hline
		L & 1.3 -- 1.6   \\
		S & 2.2 -- 2.3   \\
		C & 4.1 -- 5.1   \\
		X & 7.6 -- 8.7   \\
		U & 13.8 -- 15.6 \\
		K & 23.9 -- 24.4 \\
		Q & 43.1 -- 43.9 \\
		W & 86.2 -- 86.3 \\
		\hline
	\end{tabular}
	\label{tab:table_freqs}
\end{table}

\begin{table*}
    \centering
    \caption{Apparent angular sizes of the AGN VLBI core measured at frequencies ranging from 1.4 to 86~GHz with separate records for every epoch.}
    \label{tab:exp_data}
        \begin{tabular}{lrcccrr}
        \hline
        \hline
            Name            & $\mathrm\nu$ & Epoch      & $\theta_\mathrm{core}$ & $\theta_{\mathrm{core}}^\mathrm{err}$  & $b$        & $l$\\
                            & (GHz)        &            & (mas)                  & (mas)                  & (deg) & (deg)\\
            (1)             & (2)          & (3)        & (4)                    & (5)                    & (6)        & (7)\\
            \hline
             J0006$-$0623   & 1.40         & 2010-08-23 & 2.601                  & 0.007                 & $-66.65$  & 93.51    \\
             J0825$+$0831	& 2.31         & 2005-07-20 & 1.483                  & 0.074                 & 24.66     & 215.92   \\
             J1842$+$7946	& 4.36         & 2020-12-13 & 0.945                  & 0.098                 & 27.07	 & 111.44 \\
             J0158$+$2124   & 8.68         & 2014-08-09 & 0.686                  & 0.006                 & $-38.86$  & 142.95   \\
             J1337$-$1257   & 15.4         & 1995-07-28 & 0.091                  & 0.010                 & 48.37     & 320.02   \\
             J1048$-$1909 & 24.4          & 2002-08-25 & 0.057                  & 0.003 
             & 34.91 & 266.75\\
             J2258$-$2758	& 43.1         & 2003-09-13 & 0.089                  & 0.002                 & $-64.92$  & 24.39   \\
             J0136$+$4751	& 86.3         & 2002-04-20 & 0.028                  & 0.001                 & $-14.32$  & 130.79    \\
        \hline
        \end{tabular}
        \begin{tablenotes}
            \item The columns are as follows: (1) source name in the J2000.0 notation; (2) central observing frequency; (3) epoch; (4) measured AGN core size; (5) formal model depended error of the source size fitting; (6) Galactic latitude; (7) Galactic longitude.
            The table is published in its entirety in the machine-readable format as supplementary material. Eight randomly selected records are shown here for guidance regarding its form and content.
        \end{tablenotes}
\end{table*}

\section{Angular broadening of the AGN core size}
\label{ch:angular_broadening}

To study the scattering properties of the interstellar medium in our Galaxy, we have analyzed the VLBI core size measurements described in \autoref{ch:size_measurements}. There are 61\,230 measurements of 8\,959 active galactic nuclei which pass our filtering criterion. These AGNs are located over the entire sky, except for the region of far southern latitudes of the celestial sphere due to the lack of robust observational data in this area.

\begin{figure*}
    \centering
    \includegraphics[width=0.45\linewidth]{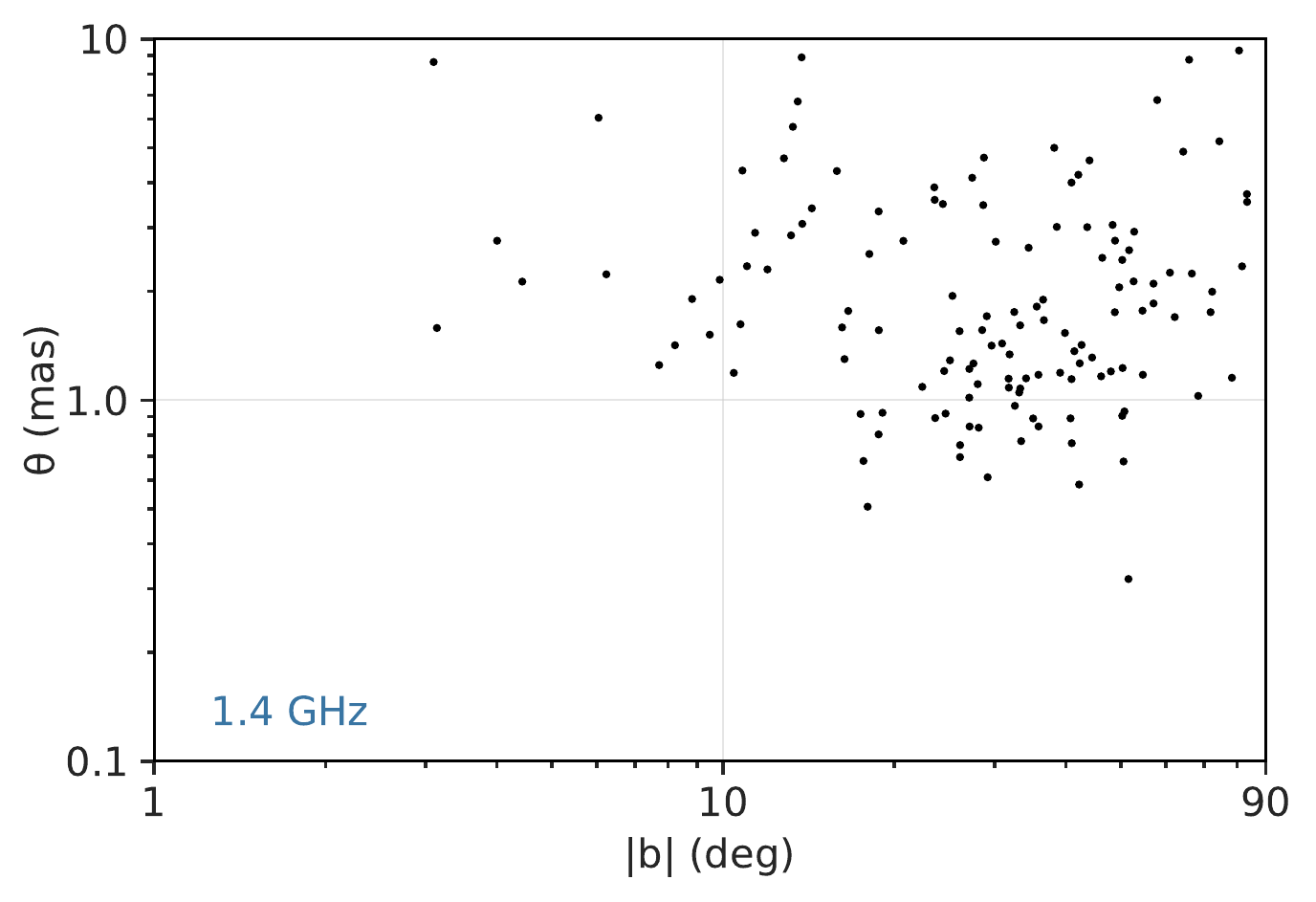}
    \includegraphics[width=0.45\linewidth]{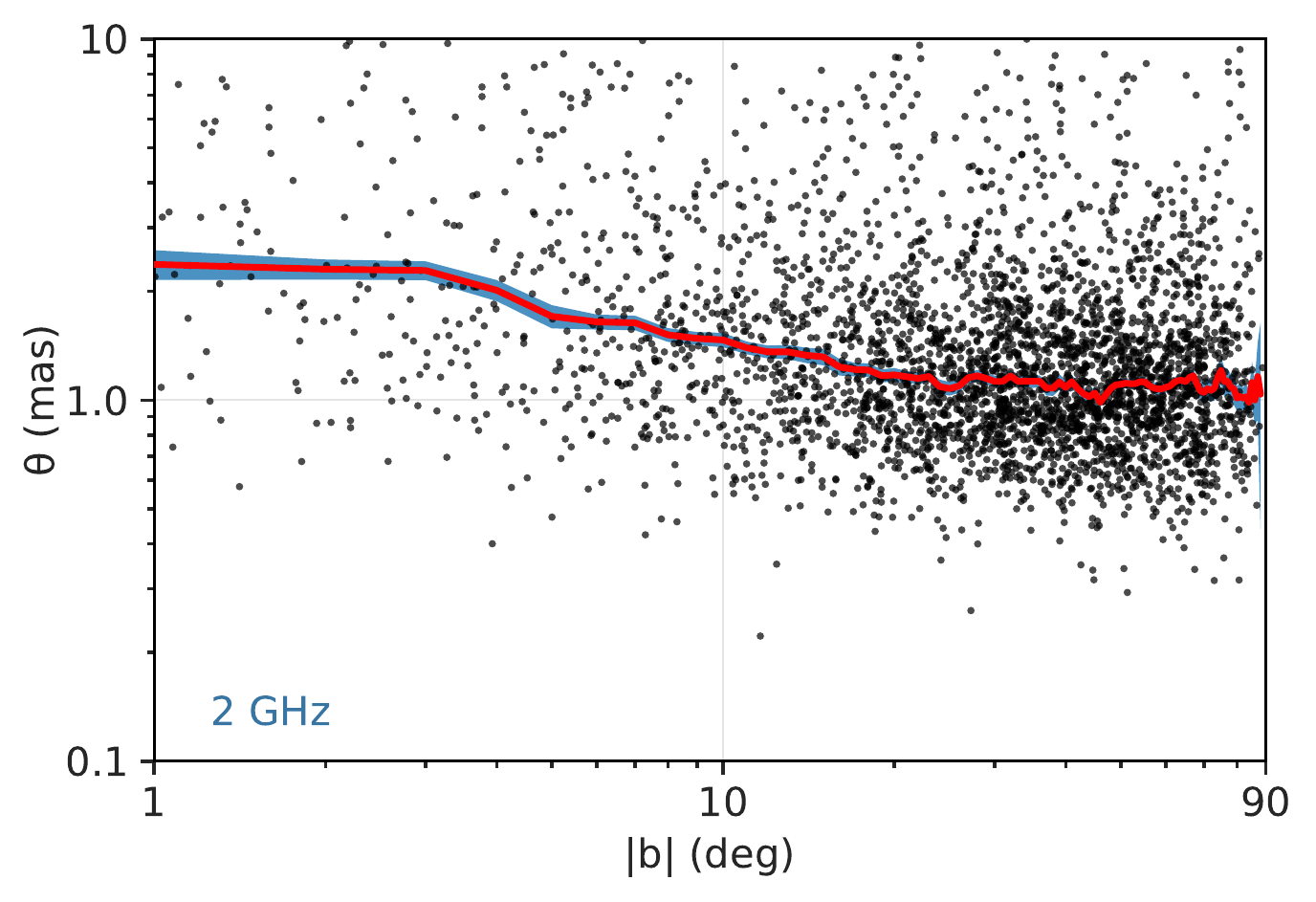}
    \includegraphics[width=0.45\linewidth]{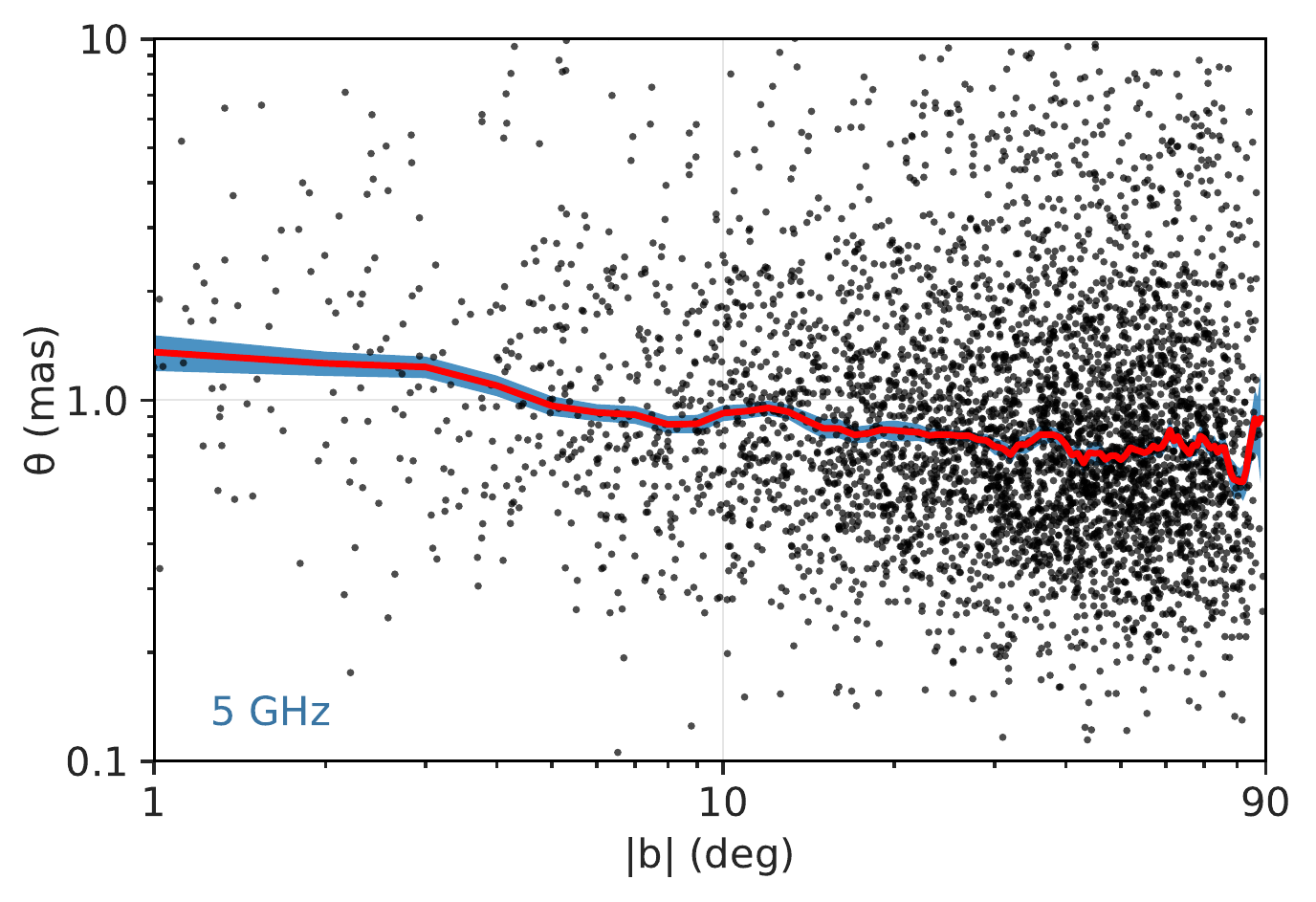}
    \includegraphics[width=0.45\linewidth]{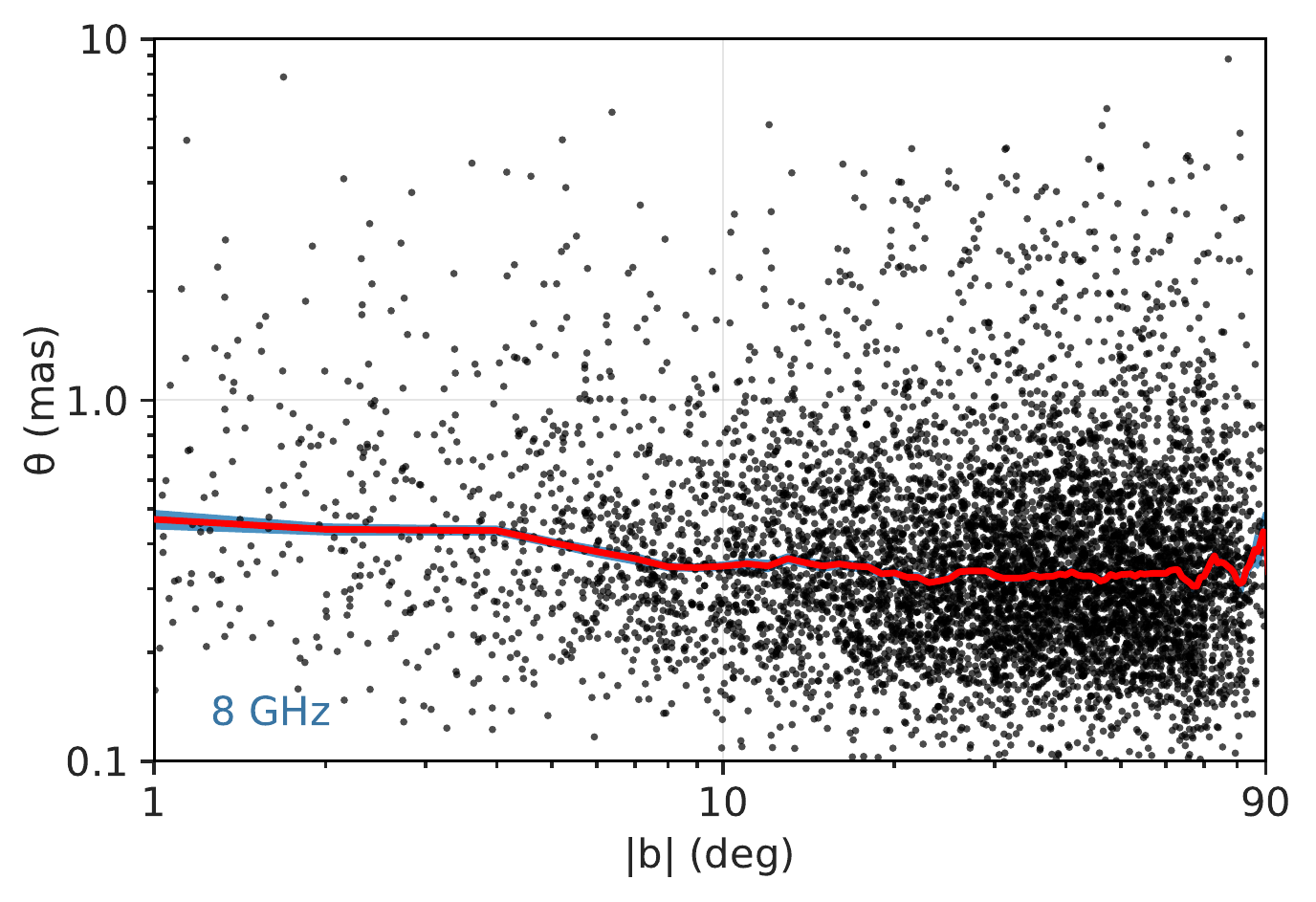}
    \includegraphics[width=0.45\linewidth]{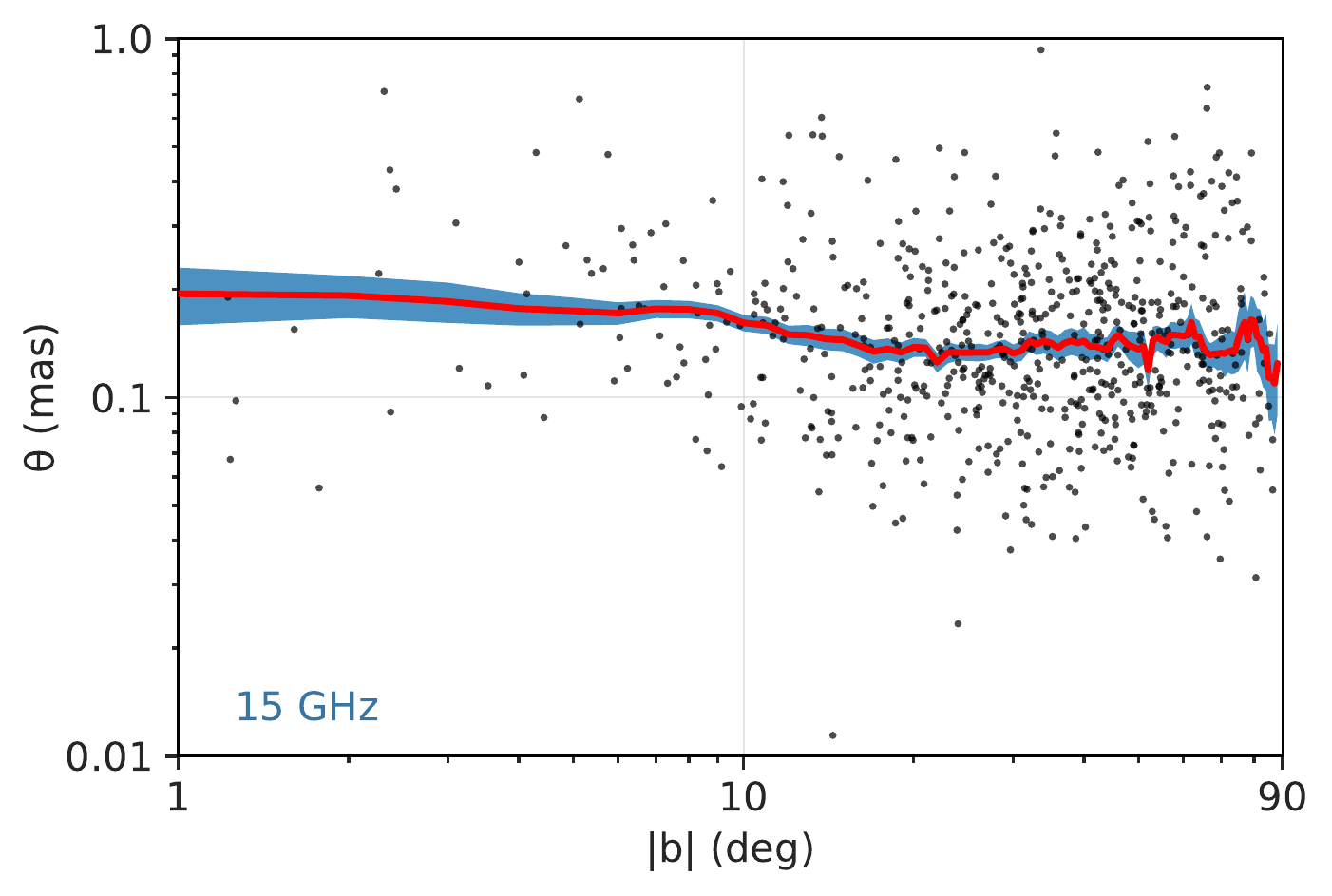}
    \includegraphics[width=0.45\linewidth]{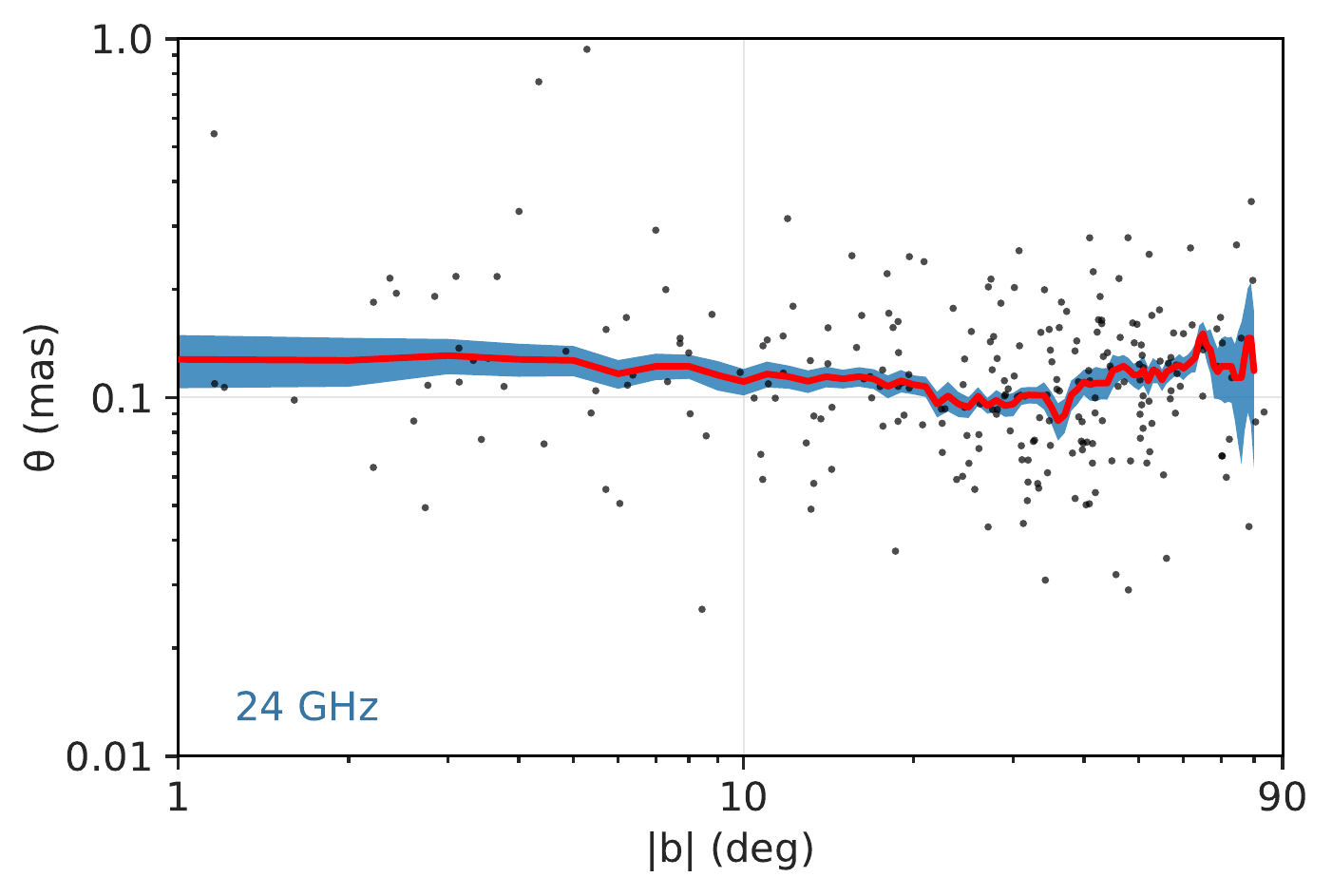}
    \includegraphics[width=0.45\linewidth]{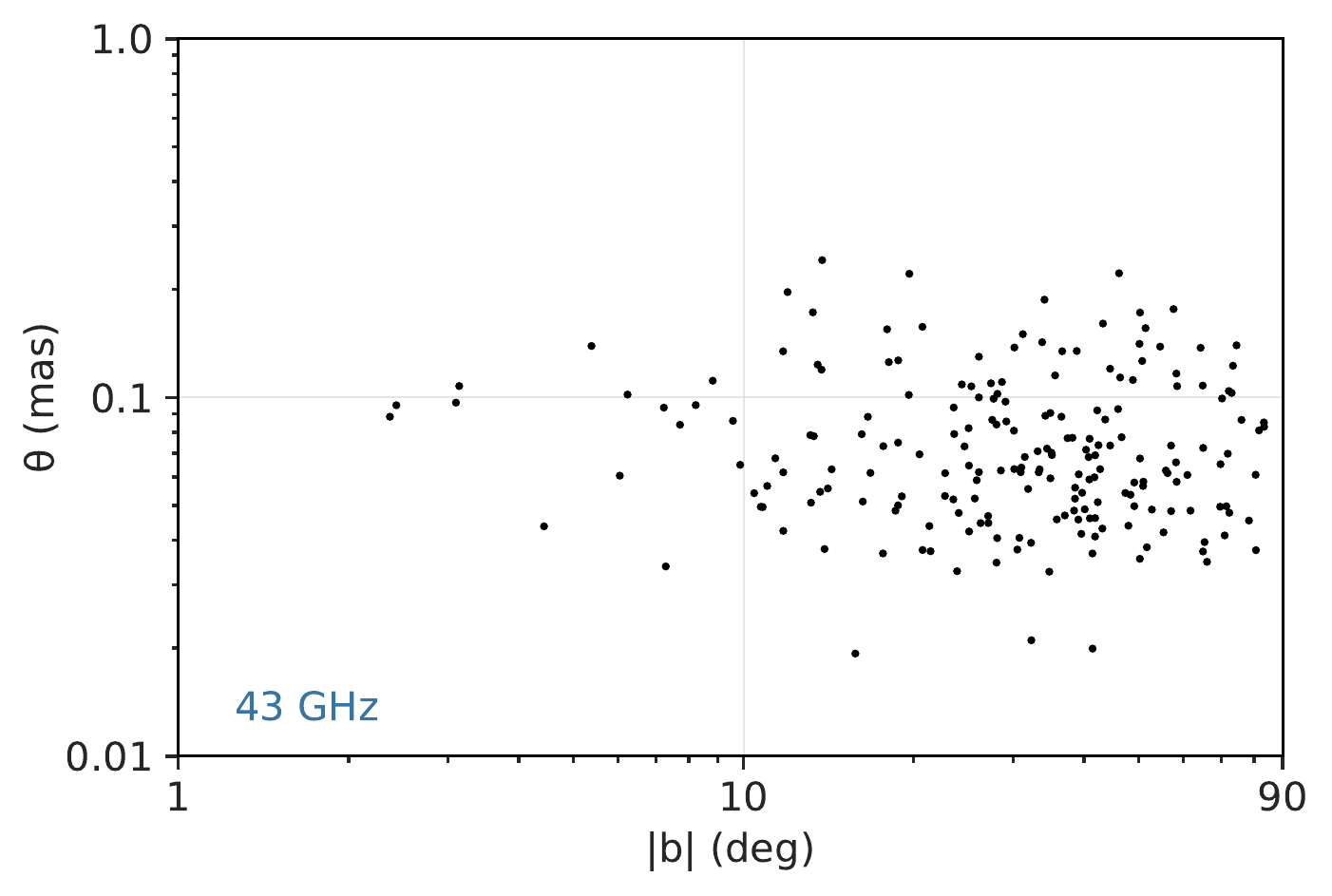}
    \includegraphics[width=0.45\linewidth]{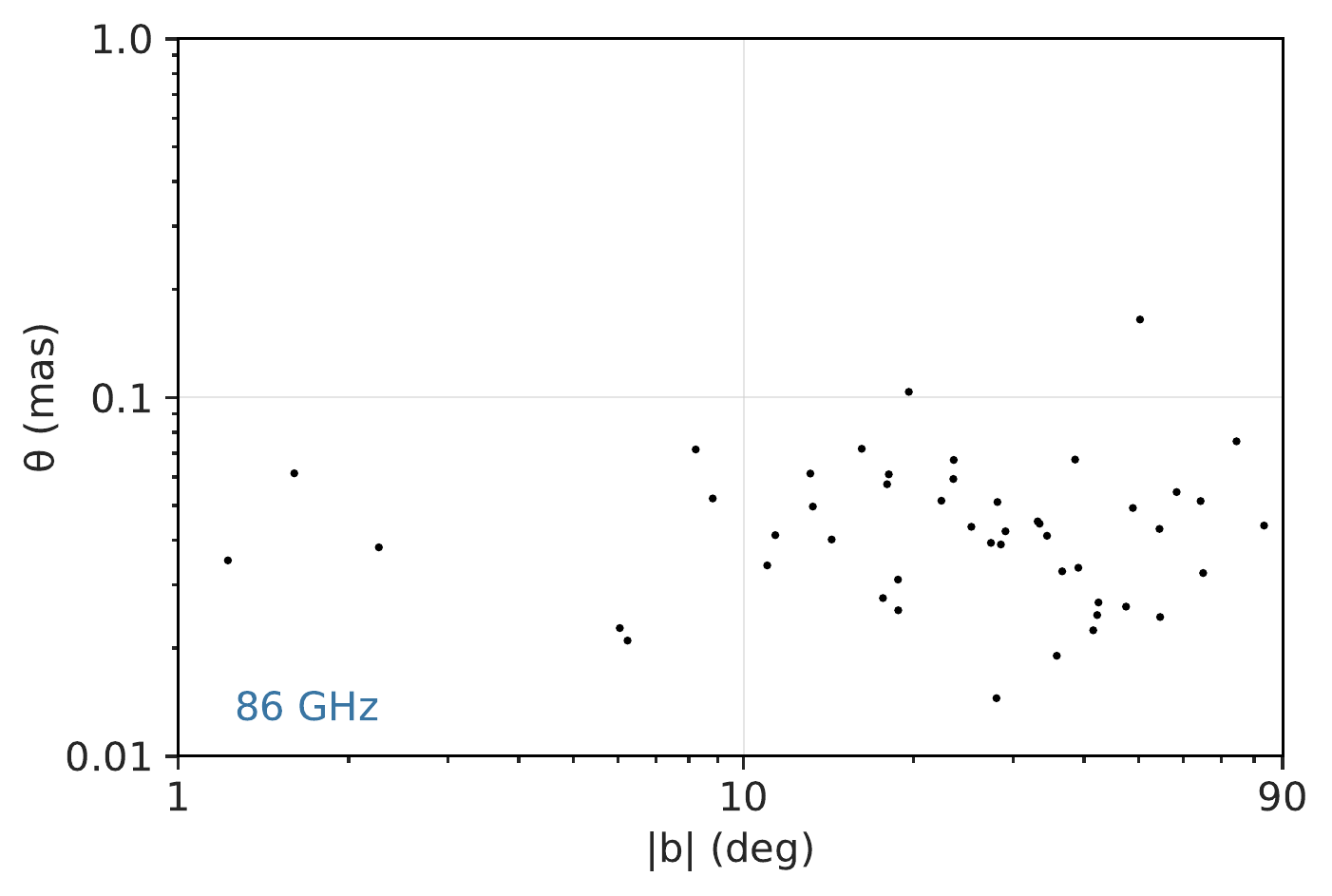}
    \caption{The AGN VLBI core sizes as a function of the absolute value of the Galactic latitude measured at frequencies ranging from 1.4~GHz to 86~GHz. Each dot represents an individual source for which the median size for all epochs was taken. The red curve is the running median which demonstrates an increase of the median sizes of the AGN cores as they approach the Galactic plane. The running median was taken over a range of $5^\circ$ of $|b|$ at 2\,--\,8 GHz and $10^\circ$ for 15~GHz, 24~GHz. The blue shaded area shows the standard deviation of the median value of the angular size.}
    \label{fig:size_vs_gallat}
\end{figure*}

Electron density fluctuations in the ISM lead to angular broadening of radio sources. Scattering material spreads throughout the Galaxy, but it mostly concentrates in a thin disc of about 100~pc in width, likely associated with HII regions, stellar wind bubbles and supernova explosions \citep{Geldzahler1981, Cordes1984, spangler1986interstellar}. \autoref{fig:size_vs_gallat} shows the AGN core sizes as a function of the absolute value of the Galactic latitude $|b|$ at all observing frequencies between 1.4~GHz and 86~GHz. These figures demonstrate that the median size of cores increases significantly when a source is seen through the Galactic plane ($|b| < 10^\circ$). This effect is most noticeable at low frequencies 2, 5, or 8~GHz, being most sensitive to scattering of radio wave and, accordingly, to angular broadening of the observed AGN core size. 

\begin{figure*}
	\includegraphics[scale=1.54]{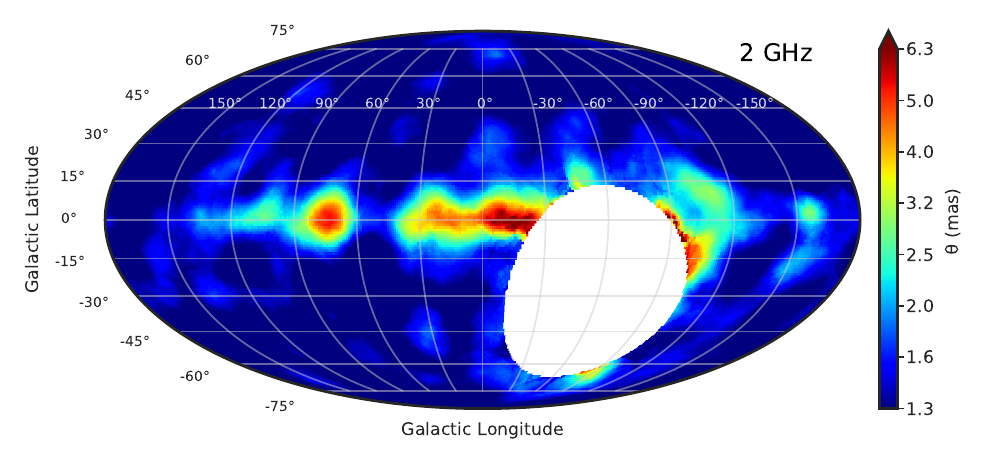}
	\includegraphics[scale=1.54]{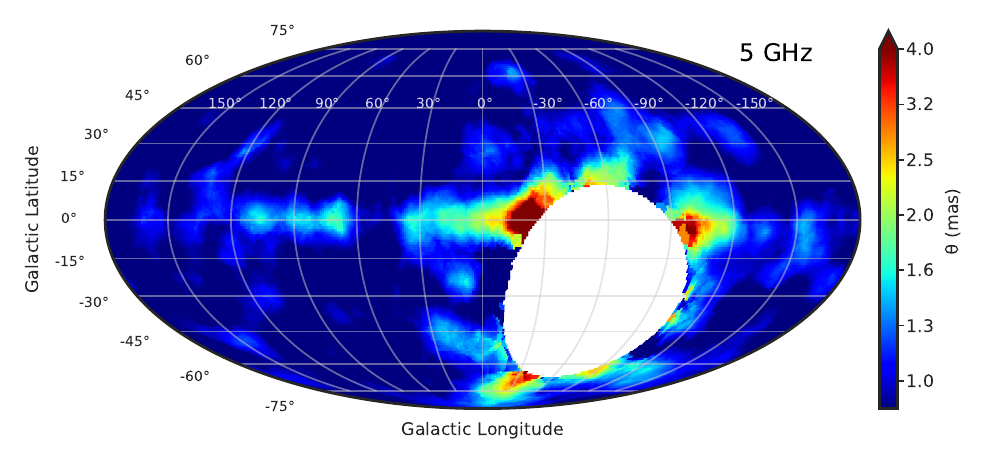}
	\includegraphics[scale=1.54]{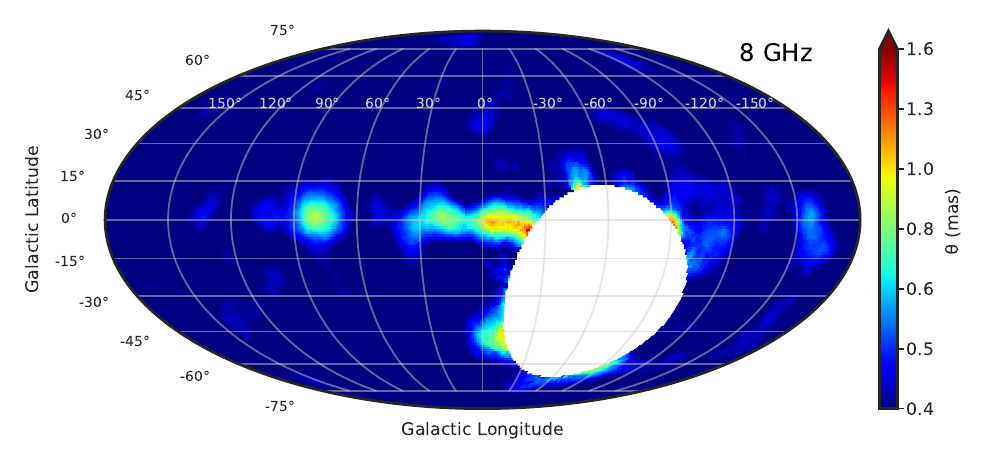}
	
    \caption{The observed AGN core size distribution maps. The colour of each pixel reflects the average size of sources at this location. The average size was estimated based on the sources which fall into a circular area of $10^\circ$ radius around this central point (pixel). The red colour corresponds to the larger observed AGN core sizes (dominance of scattering), and the dark blue colour means the smaller values of the observed AGN core sizes (weak or no scattering). All the spheres presented in this work are shown in the Galactic coordinates over the celestial sphere in the Mollweide equal-area projection. The maps are published in the FITS format as supplementary material.}
    \label{fig:size_colormaps}
\end{figure*}

The isotropic distribution of the investigated sources over the sky allowed us to map out a sky distribution of the source sizes at different frequencies to visualise the angular broadening effect (\autoref{fig:size_colormaps}). We measured the angular diameter for 3541, 4544, and 7039 AGN cores at 2, 5, 8~GHz, respectively. At other frequencies, the number of sources is considerably smaller. To create these maps, we excluded all the sources with core sizes greater than 20 mas for 2~GHz data, greater than 15~mas for 5~GHz data and greater than 10~mas for 8 GHz~data. These sources with very large sizes introduce a lot of noise to the resulting map. Each $1^\circ\times1^\circ$ pixel of the map contains the average (Gaussian function weighted) observed core size of AGN which falls into a circular area of $10^\circ$ radius around a given pixel. The empty area around the South pole is the region where we do not have enough observational data yet.

The distributions of AGN core sizes over the sky for the three frequencies are shown in \autoref{fig:size_colormaps}. A characteristic increase of the average value of measured sizes in the Galactic plane for all frequencies is observed. It is also notable that the average sizes of the sources in the Galactic plane at lower frequencies are much larger, (e.g., the map for 2~GHz, \autoref{fig:size_colormaps}, top panel) than at higher frequencies (e.g., 8~GHz, \autoref{fig:size_colormaps}, bottom panel). This effect is a result of a power-law frequency dependence of the observed AGN core size $\theta\propto\nu^{-k}$, where $k$ is expected to be 1 for the sources with undetected scattering \citep{BK79,Konigl1981} and close to 2 for the scattered ones \citep{Cordes86,Rickett1990,Armstrong95}. \autoref{tab:median_sizes} shows the median sizes of sources in the Galactic plane and outside of it derived for different frequencies. The difference between the median size values is especially noticeable at low frequencies, while at high frequencies there is almost no difference between the median sizes within and outside the Galactic plane.

\begin{table}
	\caption{Median sizes of the measured AGN VLBI cores as presented at \autoref{fig:size_vs_gallat}.}
	\centering
	\begin{tabular}{rcrcr} 
		\hline
		\hline
		$\nu$ & $\theta_\mathrm{med}\,\,(|b|<10^\circ)$ & $N$ & $\theta_\mathrm{med}\, (|b|>10^\circ)$ & $N$\\
		(GHz) & (mas)  & & (mas) & \\
	    (1) & (2) & (3) & (4) & (5)\\
		\hline
		1.4  & $2.142\pm0.895$   & 14 & $1.666\pm0.132$ & 121\\
		2    & $1.838\pm0.084$   & 510 & $1.132\pm0.012$ & 3031\\
		5    & $1.036\pm0.050$   & 544 & $0.764\pm0.012$ & 4000\\
		8    & $0.394\pm0.011$   & 979 & $0.331\pm0.003$ & 6060\\
		15   & $0.174\pm0.016$   & 64 & $0.139\pm0.004$ & 606\\
		24   & $0.127\pm0.013$   & 49 & $0.107\pm0.004$ & 210\\
		43   & $0.094\pm0.006$   & 15 & $0.064\pm0.003$ & 190\\
		86   & $0.038\pm0.012$   & 7 & $0.043\pm0.003$ & 41\\
		\hline
	\end{tabular}
	\begin{tablenotes}
       \item The columns are as follows: (1) frequency band, see \autoref{tab:table_freqs} for details; (2) median AGN core size within the Galactic plane and its error estimated with the bootstarp method; (3) number of sources used to estimate (2); (4) median AGN core size outside the Galactic plane and its error estimated with the bootstarp method; (5) number of sources used to estimate (4).
    \end{tablenotes}
    \label{tab:median_sizes}	
\end{table}

The red color on the map (\autoref{fig:size_colormaps}) corresponds to the largest observed sizes of AGN cores with the exception of those that we removed to smooth and reduce map noise, as we mentioned earlier. These sources are concentrated within a narrow band in the center of the map, where the Galactic longitude ($l$) is in a range of $-120^\circ~<~l~<~120^\circ$ for 2~GHz. The Galactic centre region is located approximately in the range of longitudes of $-20^\circ\leqslant~l \leqslant~20^\circ$. As expected, the smallest number of sources with large observed sizes is in the regions of longitudes $l~>~150^\circ$ and $l~<~-150^\circ$ --- this is the Galactic anti-centre, which contains much less scattering material. It is also clearly seen that the sources at high Galactic latitudes ($|b|>10^\circ$) are not subject to significant angular broadening. As it was demonstrated, the sources with small angular sizes fill almost the entire sphere, with the exception of some regions in the Galactic plane. This means that there are certain Galactic regions which cause strong scattering at this frequency range.

Interestingly, there are indications of scattering near the edge of the empty region. As we will see later in \autoref{ch:two_freqs_method}, the data near the edge have poorer accuracy due to the following reasons: (i) a smaller number of sources, see \autoref{fig:sphere_2_and_8_GHz}, and (ii) the angular resolution is lower because of a larger restoring beam size for the low-declination sources.
We compared the sky distribution of the local regions with cloud interaction presented in Figure~2 of \cite{Linsky2008} with the scattering maps obtained in our work, for example, the core size distribution at 2~GHz (\autoref{fig:size_colormaps}, top). There are some correspondences in the location of the scattering regions, e.g, the screen around the Cygnus supernova remnant and the region in the Galactic plane near the Galactic longitude of about $-90^\circ$. However, we do not observe a full correspondence of the detected regions of significant or possible scattering, as our data are sensitive not only to the local but also more distant screens in the Galaxy.

\begin{figure}
    \centering
	\includegraphics[width=0.95\linewidth]{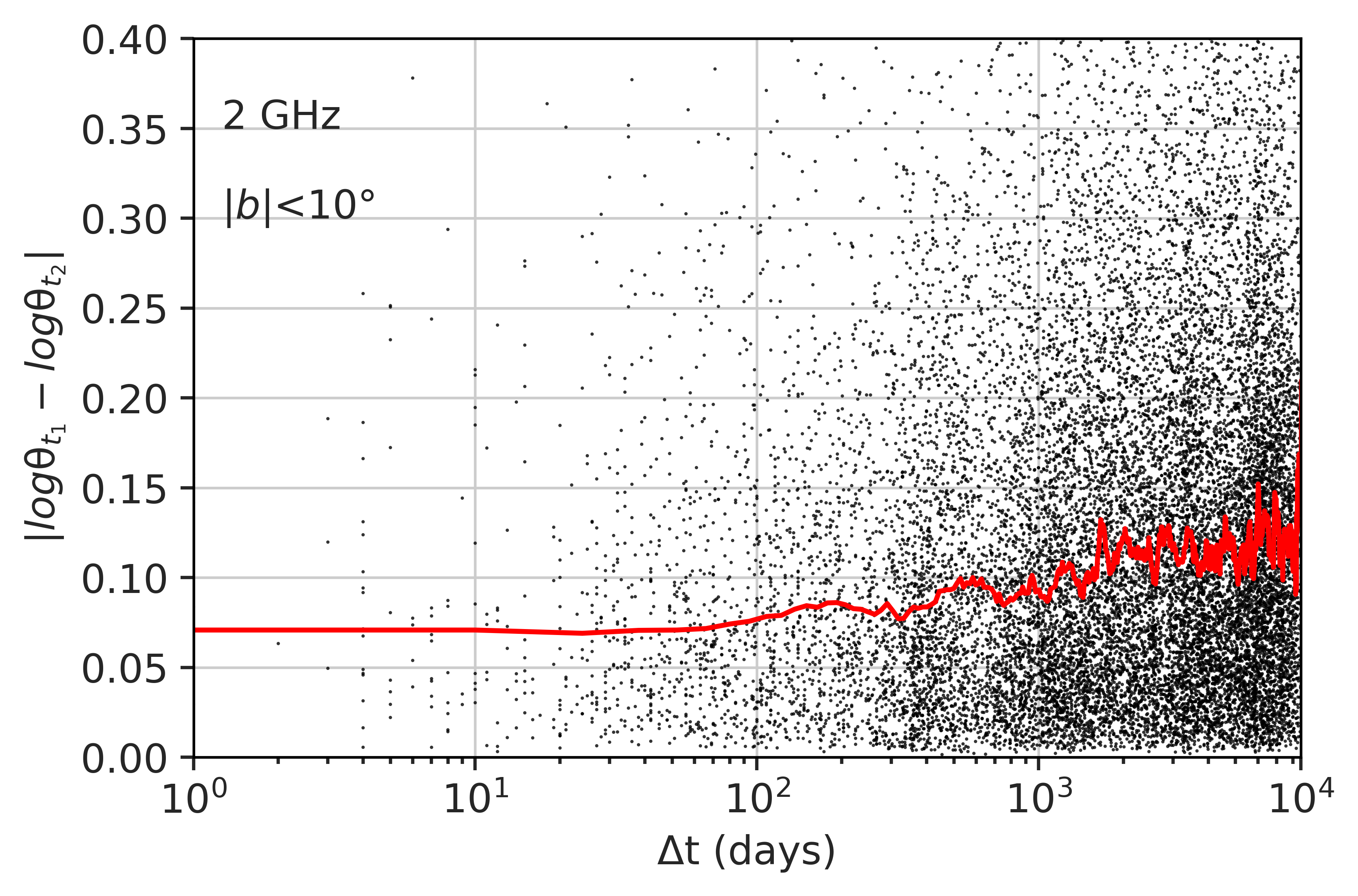}
	\includegraphics[width=0.95\linewidth]{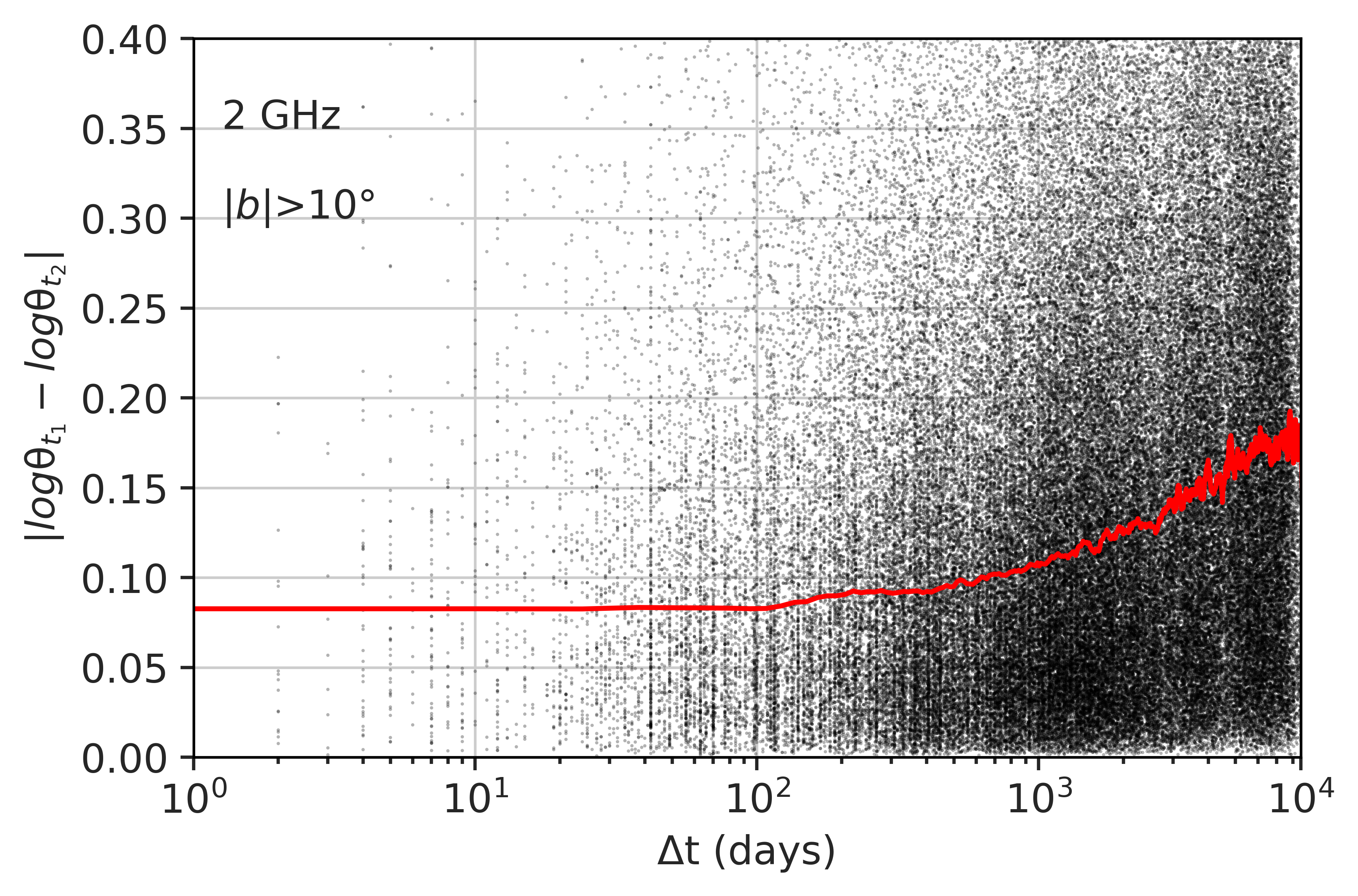}
    \caption{Amplitude of the variability of the AGN core size at 2~GHz over a wide range of time covered by observations (days to tens of year). Each dot in these plots represents the relative amplitude of a change of the AGN core sizes at 2~GHz. The running median (the red curve) was taken over a range of 100 days.}
    \label{fig:variability}
\end{figure}

It is widely known that a distinctive feature of AGNs is their high intrinsic variability across the whole electromagnetic spectrum on scales from hours to years \citep[e.g.,][]{1997ARA&A..35..607Z,2019ARA&A..57..467B,2019MNRAS.485.1822P}. To search for the characteristic time scales of the external variability of scattering properties, we used our measured multi-epoch data of the AGN VLBI core sizes at 2~GHz and 8~GHz. We calculated the size variability at a given frequency, which exceeds the variability allowed by the estimated error. In \autoref{fig:variability} we show the size variability at 2~GHz. Namely, the difference between two measured source sizes at all available time intervals $\Delta t$. All possible time periods were used for each source. Approximately 8~per~cent of all calculated amplitudes at 2~GHz and 6~per~cent at 8~GHz were excluded from the analysis, because they do not exceed the estimated error for this amplitude. 
Results can be summarized as follows.
We observe the same general properties of the core sizes at 2~GHz and 8~GHz: 
(i) variability on long time scales is much higher than on short time scales; 
(ii) variability outside the Galactic plane is higher than the variability within the plane for $\Delta t$ more than a few years, reflecting time scales of the intrinsic changes of the core size; 
(iii) on average, variability does not exceed 20~per~cent for timescales more than a year; for time scales less than a year, median amplitude remains constant. This indicates that the observed variability is dominated by internal AGN evolution rather than scattering. Thus, we did not manage to reveal the characteristic timescale of variability of scattering in the Galaxy due to the lack of data at close epochs and dominating internal AGN variability.

\section{Single power-low approximation of the $k$-index}
\label{ch:two_freqs_method}

The diffraction phenomena associated with the radiation scattering in a turbulent interstellar medium cause the angular broadening of a distant background radio source. Thus, the observed distribution of the source brightness is a convolution of the intrinsic structure of the source with the scattering function. The measured angular diameter of the AGN core should correspond to a convolution of the intrinsic and scattered components of the size \citep{Lazio08}:
\begin{gather}
    \theta_\mathrm{obs}^2 =\theta_\mathrm{int}^2 + \theta_\mathrm{scat}^2 
	\label{eq:formula1},
\end{gather}  
\begin{gather}
    \theta_{\mathrm{int}} = \theta_\mathrm{int1} \cdot \nu^{-k_{\mathrm{int}}}
	\label{eq:formula1_1},
\end{gather}
\begin{gather}
    \theta_{\mathrm{scat}} = \theta_\mathrm{scat1} \cdot \nu^{-k_\mathrm{scat}}
	\label{eq:formula1_2},
\end{gather}
where $\theta_\mathrm{obs}$ is the measured AGN core size at the observing frequency, $\theta_\mathrm{int}$ and $\theta_\mathrm{scat}$ are the intrinsic and scattered AGN core sizes at the observing frequency, $\theta_\mathrm{int1}$ and $\theta_\mathrm{scat1}$ are the intrinsic and scattered AGN core sizes at 1~GHz, $\nu$ is the frequency of observation (GHz), $k_{\mathrm{int}}$ is the power-law index from the frequency dependence of the intrinsic core size and $k_{\mathrm{scat}}$ is the power-law index from the frequency dependence of the scattered AGN core size.

\begin{figure}
	\includegraphics[width=\linewidth]{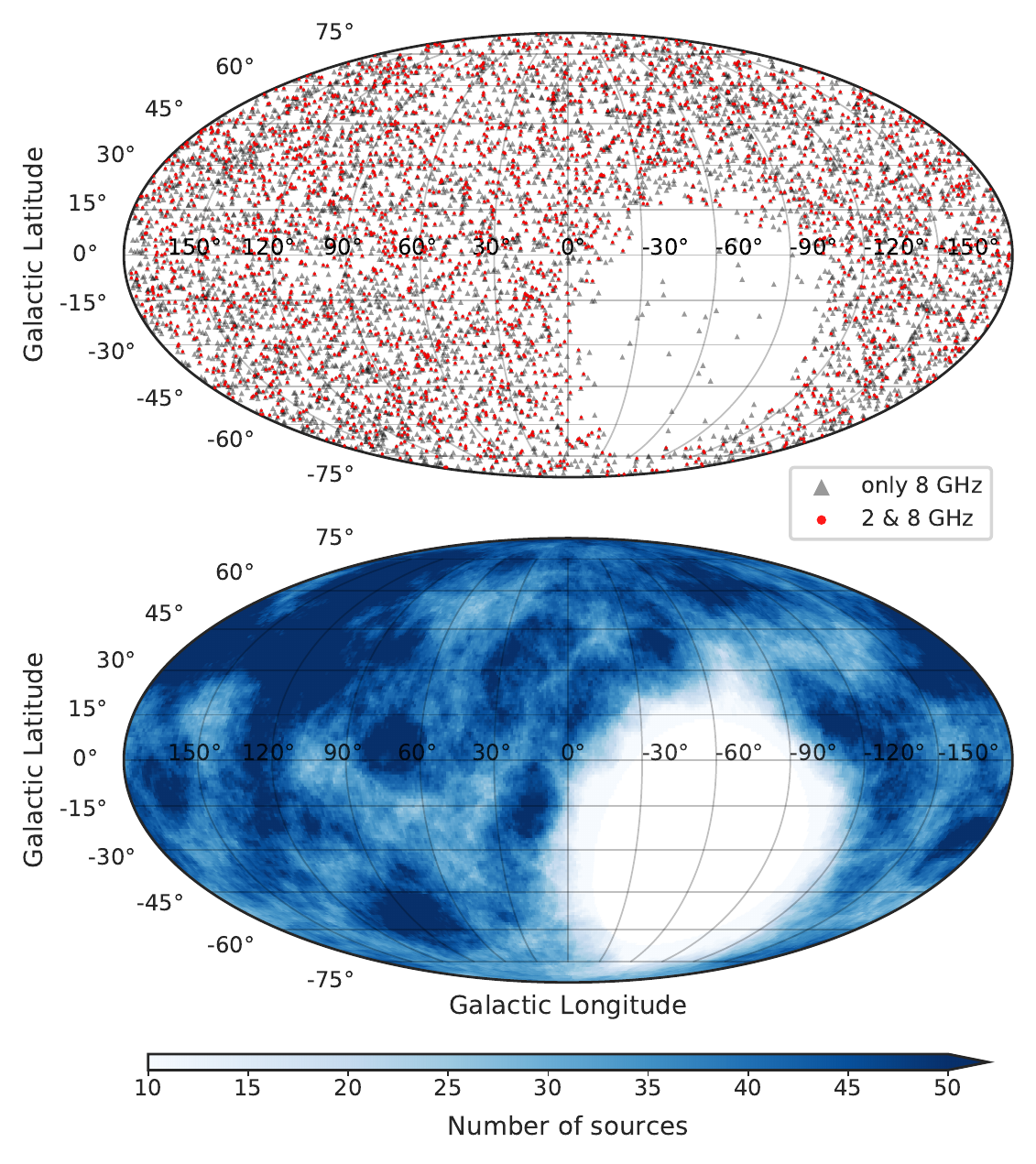}
    \caption{Top panel: Sky distribution of the measured AGN core sizes. The red dots represent 2614 AGNs for which the angular size was simultaneously measured at 2~GHz and 8~GHz. The grey triangles are 7039 AGNs measured only at 8~GHz. The bottom panel shows the averaged density distribution map for simultaneously measured AGN core sizes at 2~GHz and 8~GHz. Each pixel of the map reflects the number of the sources which fall into a circular area of $10^\circ$ radius around this central point (pixel).}
    \label{fig:sphere_2_and_8_GHz}
\end{figure}

If we consider a simple approximation to Equations~\ref{eq:formula1}--\ref{eq:formula1_2}, it can be expressed as $\theta_{\mathrm{obs}}\propto\nu^{-k}$. It is expected that the index $k$ is approximately equal to 2, if scattering dominates and is equal to 1 when the radiation coming from a source is not scattered. This power-law $k$ index reflects the strength of scattering of the source and will allow us to analyse its distribution over the sky, similar to the distribution of the observed AGN core sizes at different frequencies, which we showed in \autoref{fig:size_colormaps}.

To estimate the index $k$, we can use the data of simultaneous multi-epoch measurements of the AGN core sizes at two frequencies, 2~GHz and 8~GHz. We calculated the index $k$ for 2614 AGNs. If a source has more than one observation epoch, the median value of the $k$ index was used. \autoref{fig:sphere_2_and_8_GHz} (top panel) demonstrates the sky distribution of these sources (red dots). We also plotted in \autoref{fig:sphere_2_and_8_GHz} (top panel) the sources measured only at 8~GHz (gray triangles). \autoref{fig:sphere_2_and_8_GHz} (bottom panel) also shows a source density distribution map based on the data from simultaneous observations at 2~GHz and 8~GHz. Each pixel of the map reflects the average number of sources that fall into a circular area of $10^\circ$ radius around this central pixel.

\begin{figure*}
    \centering
    \includegraphics[width=0.48\linewidth]{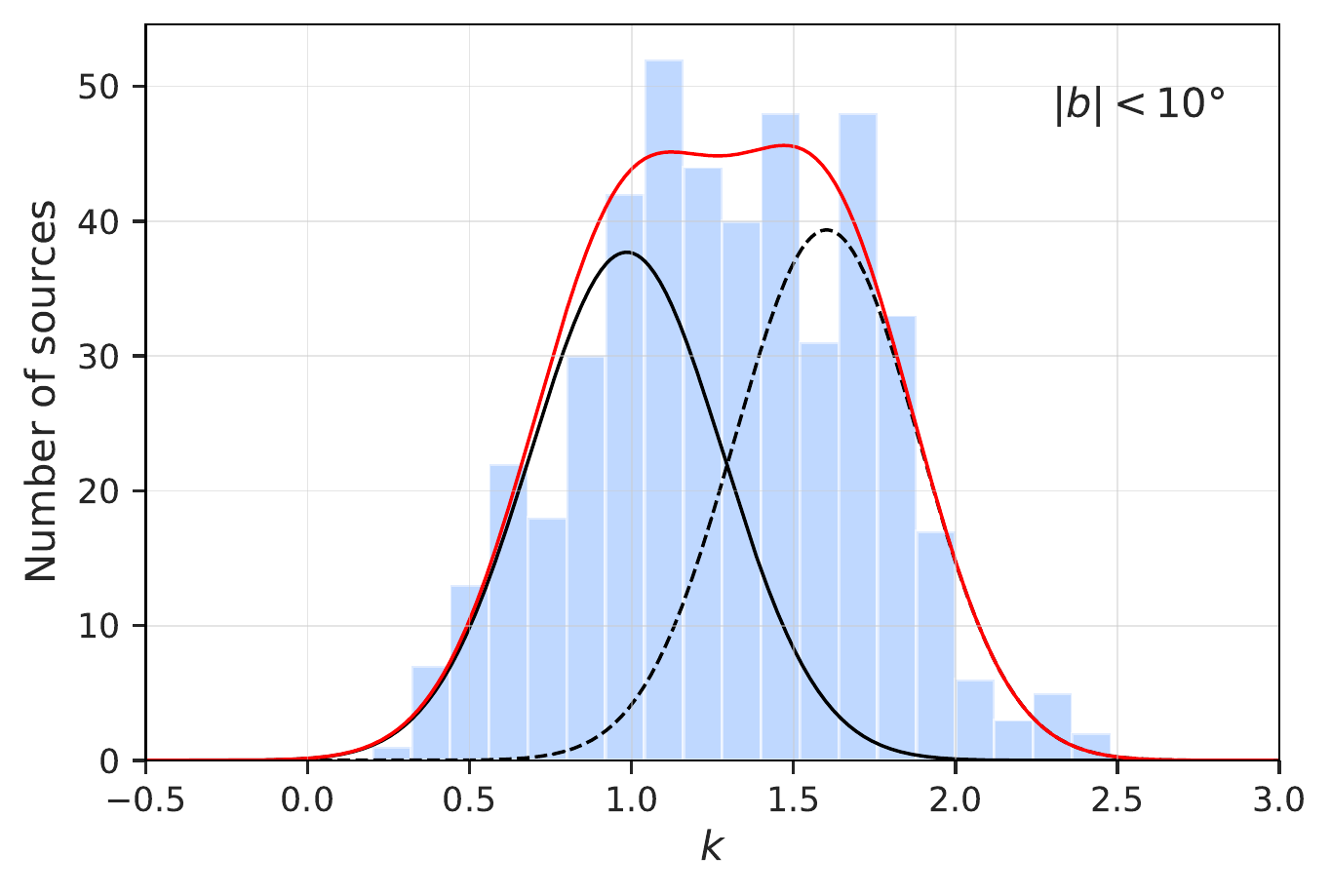}
    \includegraphics[width=0.49\linewidth]{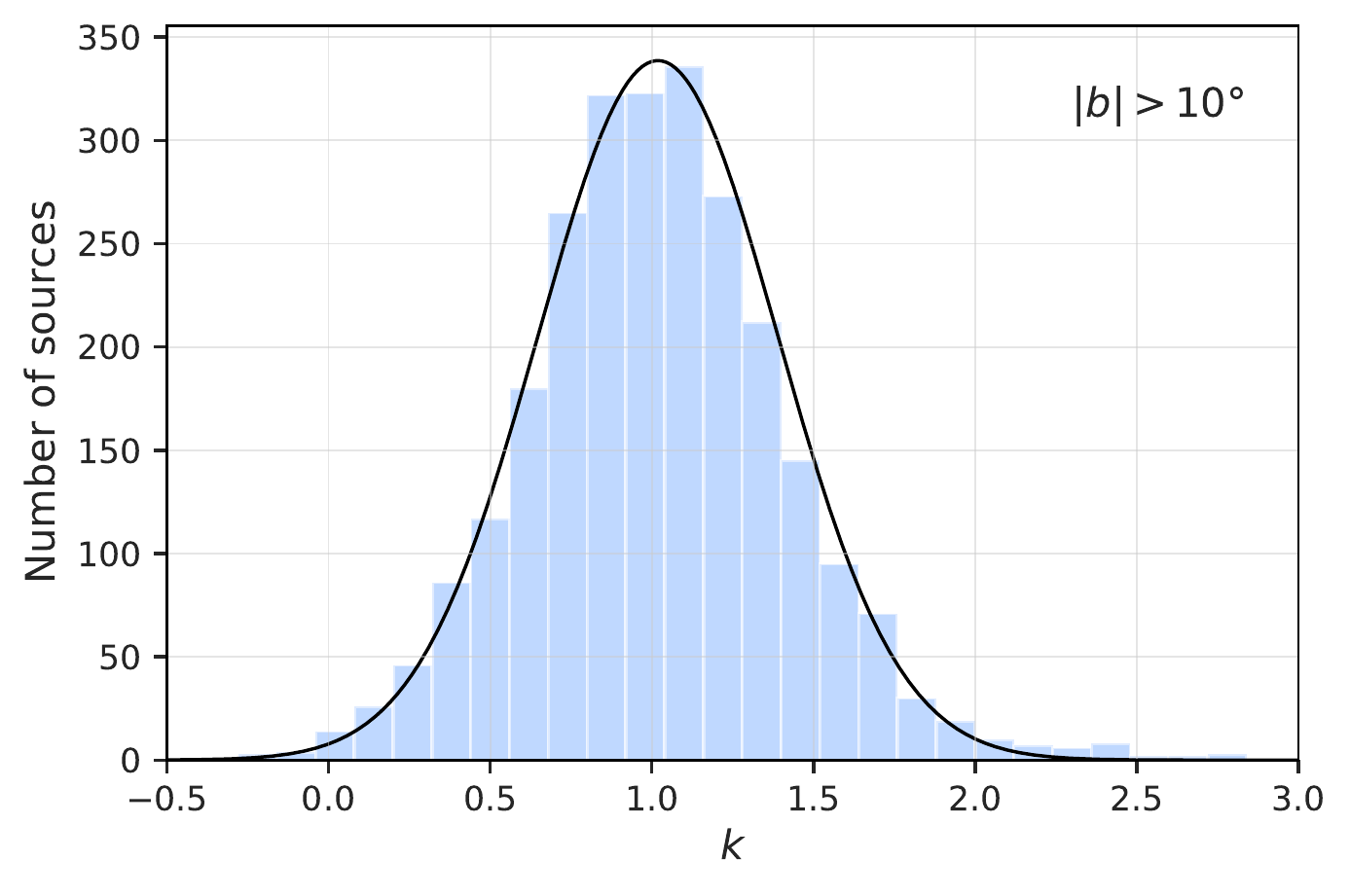}
    \caption{Histograms of the $k$ indices, derived from simultaneous measurements of the AGN core sizes $\theta\propto\nu^{-k}$ at 2~GHz and 8~GHz. The left histogram is the distribution of indices in the Galactic plane ($|b| < 10^\circ$), fitted with two Gaussian curves representing unscattered and scattered sources. The right histogram is the distribution of indices outside the Galactic plane ($|b| > 10^\circ$), fitted by a single Gaussian curve. All the parameters of the fitting functions are listed in \autoref{tab:table_gaussians}. The red curve presented on the left histogram shows the distribution formed by the sum of the first (solid) and the second (dashed) Gaussian curves.}
    \label{fig:k_28_hists}
\end{figure*}

The derived $k$ index values are shown in \autoref{fig:k_28_hists}, separately for 462 and 2614 sources seen through the Galactic plane (\autoref{fig:k_28_hists}, left) and outside of it (\autoref{fig:k_28_hists}, right), with the median $k$ values 1.29 and 1.01, respectively. The distributions of $k$ indices were fitted using the Gaussian functions: two Gaussians for data within the Galactic plane to separate contributions of scattered and unscattered sources and a single Gaussian for the region outside the Galactic plane. The obtained parameters of the Gaussian curves are presented in \autoref{tab:table_gaussians}.

\begin{figure}
	\includegraphics[width=\linewidth]{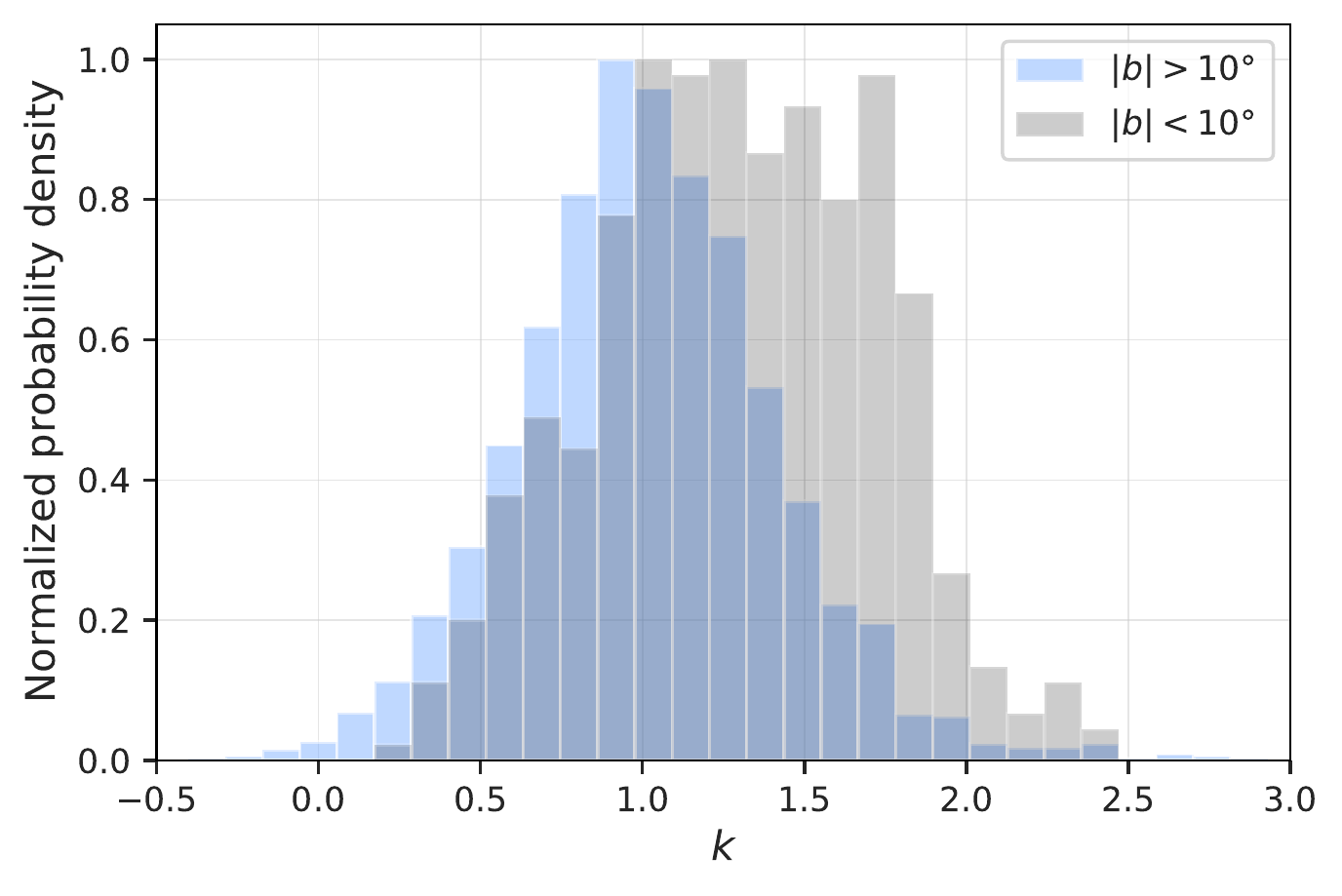}
    \caption{Amplitude-normalised histogram of the $k$ indices, derived from simultaneous measurements of the AGN core sizes at 2~GHz and 8~GHz in the Galactic plane (grey histogram) is plotted on top of the histogram of the $k$ values measured outside the Galactic plane (blue histogram).}
    \label{fig:k28_another_variant}
\end{figure}

\autoref{fig:k28_another_variant} is the amplitude-normalised histogram of the $k$ indices, where the data in the Galactic plane is plotted on top of the $k$ values measured outside the Galactic plane. That shows that the values of the $k$ index outside of the Galaxy (\autoref{fig:k_28_hists}, right) cluster around a value of $k = 1.02~\pm~0.01$ and the first Gaussian curve within the Galaxy (\autoref{fig:k_28_hists}, left) with a peak at $k = 0.99~\pm~0.02$ describe the distributions of sources showing similarly insignificant levels of scattering. Notably, the obtained peak values of the $k$ indices are in good agreement with the theoretically expected $k = 1$ for the frequency dependence $\theta_{\mathrm{obs}}\propto\nu^{-k}$ of the intrinsic AGN core sizes \citep{BK79,Konigl1981}.
Note that this value is also in a good agreement with the often observed core-shift dependence $r\propto\nu^{-1}$, see for details \citet{2011A&A...532A..38S,2014MNRAS.437.3396K}.
The second Gaussian curve on the left histogram (\autoref{fig:k_28_hists}) with a peak at $k = 1.60 \pm 0.02$ describes the contribution of the scattered sources to the distribution of the $k$ indices in the Galactic plane. All the errors here are estimated as the standard deviations of the means of samples. The resulting $k$ index value differs significantly from the theoretically expected value $k\approx2$ for the scattered sources. The index value close to $k = 2$ is observed for a small number of sources in the Galactic plane. We used a simple approximation for the frequency dependence of the observed AGN core size for these calculations. It is the main reason for underestimating this value of the $k$ index for the scattered sources. All of the obtained results of the $k$ are very similar to those obtained by an earlier study by \cite{Pushkarev15}.

The presence of two modes in the $k$ indices distribution in the Galactic plane indicates that a significant number of sources are not scattered even though lie at low latitudes, and the signal from these sources passes through the most densely populated part of our Galaxy. Thus, we can expect many scatter--free regions in the Galactic plane.

\begin{figure*}
	\includegraphics[scale=1.6]{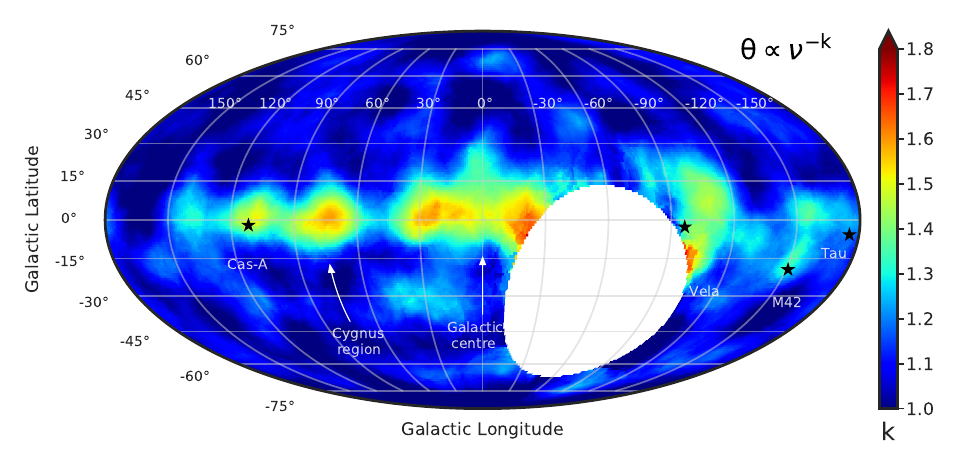}
	\includegraphics[scale=1.6]{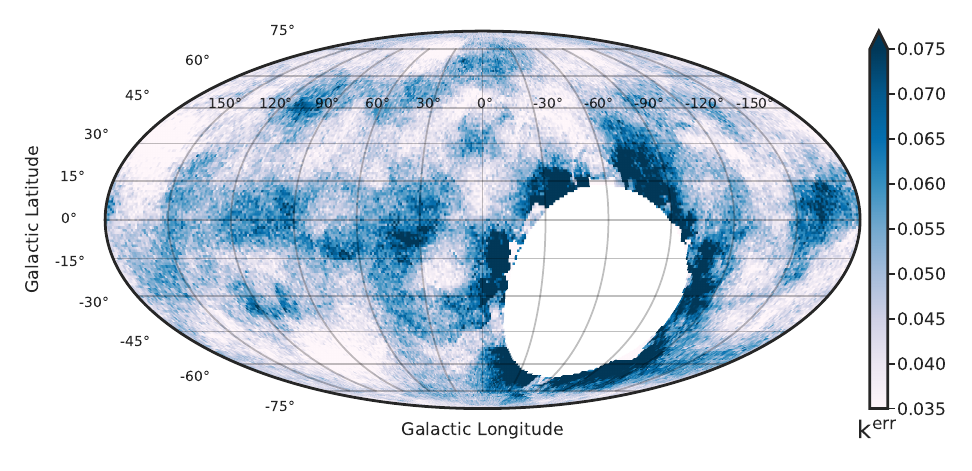}
    \caption{Top: Distribution map of the power-law $k$ index derived from the AGN core sizes $\theta\propto\nu^{-k}$ simultaneously measured at 2~GHz and 8~GHz over the sky. The magnitude and colour of each pixel of the map reflect the average over $10^\circ$ area value of the $k$ index at this location. The red colour corresponds to higher values of the $k$ index (dominance of scattering), the dark blue colour corresponds to the lower values of the $k$ index (the sources with undetected scattering). Bottom: The distribution map of the standard deviation of the estimated average index $k$. The darker the map area, the greater the error of the estimated average value of the $k$ index in this direction of the sky. The maps are published in the FITS format as supplementary material.}
    \label{fig:k28_map}
\end{figure*}

Using the obtained values of the $k$ indices calculated for 2614 sources uniformly distributed over the sky, we created the distribution map of scattering power in the Galaxy (\autoref{fig:k28_map}). To create this map, we excluded the sources outside the interval $0.5 \leqslant k \leqslant 2.5$, because too small and too large index values introduce needless noise. The resulting $k$ index distribution map replicates the pattern of the AGN core size distribution map at low frequencies, for example, at 2~GHz (compare with \autoref{fig:size_colormaps}, top). According to the map, we can highlight the regions of strong scattering in the Galactic plane ($|b| < 10^\circ$):
\begin{enumerate}
\item the region at $-20^\circ\leqslant l\leqslant20^\circ$ encompasses the Galactic centre and Galactic bar. This central region also contains Sagittarius~A$^\ast$ hosting a supermassive black hole surrounded by a hot radio-emitting gas cloud approximately 1.8~pc in diameter \citep{Downes1971}; 
\item Cygnus supernova remnant is located at $(l, b)~=~(74\fdg0, -8\fdg6)$;
\item Cassiopeia~A is a supernova remnant located at $(l, b)~=~(111\fdg7, -2\fdg1)$;
\item Vela supernova remnant is located at $(l, b)$ = $(-96\fdg5, -2\fdg8)$;
\item Taurus A supernova remnant is located at $(l, b)~=~(-175\fdg4, -5\fdg8)$ in the constellation of Taurus;
\item and also an area of moderate scattering in a direction to the Orion Nebula M42, the closest region of massive star formation at $(l, b) = (-151\fdg0, -19\fdg4)$. \cite{Koay2019} identified the region of Orion-Eridanus superbubble as an important scattering region even at 15~GHz due to the detection of 11 radio sources of interday variability on these sightlines.
\end{enumerate}

We also derived the $k$ indices using the two-frequency method mentioned above and the data from non-simultaneous observations of the AGN core sizes at frequency pairs 2~GHz and 8~GHz (3324 sources), 2~GHz and 5~GHz (1178 sources), 5~GHz and 8~GHz (2833 sources). For each source, the median observed core size at a given frequency was taken. A common feature for all three cases is that the $k$ index distribution for the AGNs in the Galactic plane ($|b|<10^\circ$) does not have a clear two-peak structure and cannot be reliably fitted by two Gaussian curves. The distributions of $k$ for the sources outside the Galactic plane obtained from the non-simultaneous measurements look similar and almost do not differ from the distribution represented in \autoref{fig:k_28_hists}, right. The peak of all obtained histograms is concentrated around the value $k = 1$. We conclude that simultaneous measurements are required to properly address the joint case of intrinsic and external effects of the core size. For this reason, we prefer the results from the simultaneous observations at two frequencies for the further analysis in this paper.

\section{Connection between scattering properties and rotation measure, electron density, and $\mathrm{H\alpha}$ distributions in the Galaxy}
\label{ch:corellation_tests}

The strength of scattering of radio wave emission which passes through clumpy distributed plasma is related to the length-scales and amplitude of the density inhomogeneities of the medium along the line of sight. Therefore, we can expect an enhancement of the scattering strength in regions with a high concentration of free electrons. For this reason, in this section, we will test the connection between the distribution of the scattering strength and the parameters measured in our Galaxy which are related to free electrons concentration, namely the rotation measure \citep[$RM$,][]{Taylor09}, the scattered size distribution based on the NE2001 model \citep[$\theta_{\mathrm{NE2001}}$,][]{NE2001} and the distribution of the $\mathrm{H\alpha}$ radiation intensity in the Galaxy \citep[$\mathrm{H\alpha}$,][]{Finkbeiner03}. The distribution of the scattering strength in the Galaxy, as we have already noted, can be reflected by the distribution of the AGN core sizes measured at low frequency, for example, at 2~GHz (see \autoref{ch:angular_broadening}
for details), and also by the distribution of the power-law $k$ indices from the frequency dependence of the AGN core sizes (see \autoref{ch:two_freqs_method} for details). 

We used Kendall's $\tau$ non-parametric rank correlation coefficient to assess the magnitude of the correlation between the compared parameters. The errors of $\tau$ are estimated as standard deviations of the obtained by bootstrap parameter distributions. The results of Kendall's $\tau$ tests are shown in \autoref{tab:correlation_results}. The strongest correlation is traced between the AGN core sizes $\theta_\mathrm{obs,2\,GHz}$ obtained from the observations at 2~GHz and the $\mathrm{H\alpha}$ radiation intensity distribution in the Galaxy. Thus, we can conclude that there is a direct relationship between the regions of a high $\mathrm{H\alpha}$ intensity and the distribution of scattering screens in the Galaxy. The correlations of $\theta_\mathrm{obs,2\,GHz}$ with other parameters are also significant.

\begin{figure*}
	\includegraphics[scale=1.8]{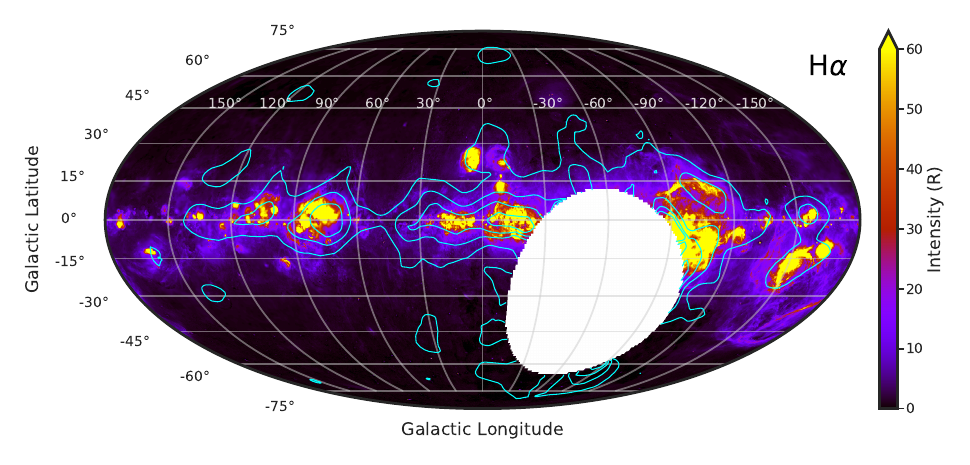}
    \caption{Distribution map of the $\mathrm{H\alpha}$ radiation intensity in the Galaxy, on top of which the observed AGN core size distribution $\log\theta_\mathrm{obs,\,2\,GHz}$~(mas) contours are plotted. The colour of the map reflects intensity in Rayleighs (1$R$ = $10^6$ photons $\mathrm{cm}^{-2}$ $\mathrm{s}^{-1}$ $\mathrm{sr}^{-1}$). The contour levels denote $\theta_\mathrm{obs,\,2\,GHz}\,=\,1.6,\,2.5,\,4.0,\,6.3\,\,\mathrm{mas}$.}
    \label{fig:theta_vs_Halpha}
\end{figure*}

\autoref{fig:theta_vs_Halpha} shows the distribution map of the $\mathrm{H\alpha}$ radiation intensity in the Galaxy \citep{Finkbeiner03}. The Finkbeiner's full sky $\mathrm{H\alpha}$ map was created using the data from several surveys: the Wisconsin H-Alpha Mapper\footnote{\url{ http://www.astro.wisc.edu/wham/}}, the Virginia Tech Spectral-Line Survey\footnote{\url{http://www1.phys.vt.edu/~halpha/}}, and the Southern H-Alpha Sky Survey Atlas\footnote{\url{http://amundsen.swarthmore.edu/SHASSA/}}. The $\mathrm{H\alpha}$ data in Cartesian projection with 6$'$ resolution are available on the D.~Finkbeiner's website\footnote{\url{https://faun.rc.fas.harvard.edu/dfink/skymaps/}}. On top of the $\mathrm{H\alpha}$ maps, we plotted contours of the average observed AGN core sizes measured at 2~GHz, $\theta_\mathrm{obs,\,2\,GHz}$. For the contours, we used the sources with the core sizes less than 20~mas to avoid unnecessary noise introduced by large individual sources. These contours largely repeat the locations of high-intensity $\mathrm{H\alpha}$ clouds. This result confirms the expectation because the $\mathrm{H\alpha}$ emission observed in spiral galaxies is a direct indicator of a hot ionised interstellar medium \citep[HII,][]{Reynolds1983}. Previously, \cite{Lovell2008} also reported on very significant correlation between scintillations caused by the turbulent interstellar medium and the line-of-sight $\mathrm{H\alpha}$ emission in the Galaxy.

We note that the correlations with $\mathrm{H\alpha}$ data may result in a higher Kendall's $\tau$ value than others due to the higher $\mathrm{H\alpha}$ data resolution. Additionally, $RM$ values partly depend on magnetic field strength, while it has no effect on scattering. In this sense, $\mathrm{H\alpha}$ turns out to be an optimal tracer of the ionized hot plasma.

\begin{table}
	\caption{Fitted parameters of the Gaussian curves for the distributions of the $k$ indices derived from the data of simultaneous observations of the AGN core sizes at 2~GHz and 8~GHz.}
	\centering
	\begin{tabular}{|*{5}{c|}} 
		\hline
		\hline
		$|b|$       & $\mu$ & $\sigma$ & $\mu_{2015}$ & $\sigma_{2015}$ \\
		(1)         &(2)    &(3)       &(4)           &(5)\\
		\hline
		$<10^\circ$ & 0.99  & 0.30     & 0.91         & 0.33\\
		            & 1.60  & 0.28     & 1.76         & 0.28\\
		$>10^\circ$ & 1.02  & 0.37     & 0.90         & 0.44\\
		\hline
	\end{tabular}
	\begin{tablenotes}
        \item The columns are as follows: (1)~$|b|$ is the absolute range of the Galactic latitude ($^\circ$); (2)~$\mu$ is the mathematical expectation obtained in this work; (3)~$\sigma$ is the standard deviation obtained in this work; (4)~$\mu_{2015}$ is the mathematical expectation obtained in \cite{Pushkarev15}; (5)~$\sigma_{2015}$ is the standard deviation obtained in \cite{Pushkarev15}.
    \end{tablenotes}
	\label{tab:table_gaussians}	
\end{table}

\section{Two-component model of the frequency-dependent AGN core size}
\label{ch:k_scatter_modeling}

This section presents Monte-Carlo simulations of the intrinsic and scattered components of the observed source size $\theta_\mathrm{obs}$. Also, we present two methods for estimating the characteristic value of the scattering index $k_\mathrm{scat}$ and then compare all the obtained results with the theoretical predictions.

\subsection{Multi-frequency fitting of observed core sizes}
\label{sec:section6_1}

Given a large amount of experimental data, we determined (\autoref{ch:two_freqs_method}) that the exponent $k$ for the sources with undetected scattering equals 1 with a high accuracy. Therefore, we can set $k_\mathrm{int}$ in Equations~\ref{eq:formula1}--\ref{eq:formula1_2} equal to 1 and determine the unknown parameters of these equations for each selected source, namely $k_\mathrm{scat}$, $\theta _\mathrm{int1}$, $\theta_\mathrm{scat1}$.

\begin{figure}
    \centering
	\includegraphics[width=\linewidth]{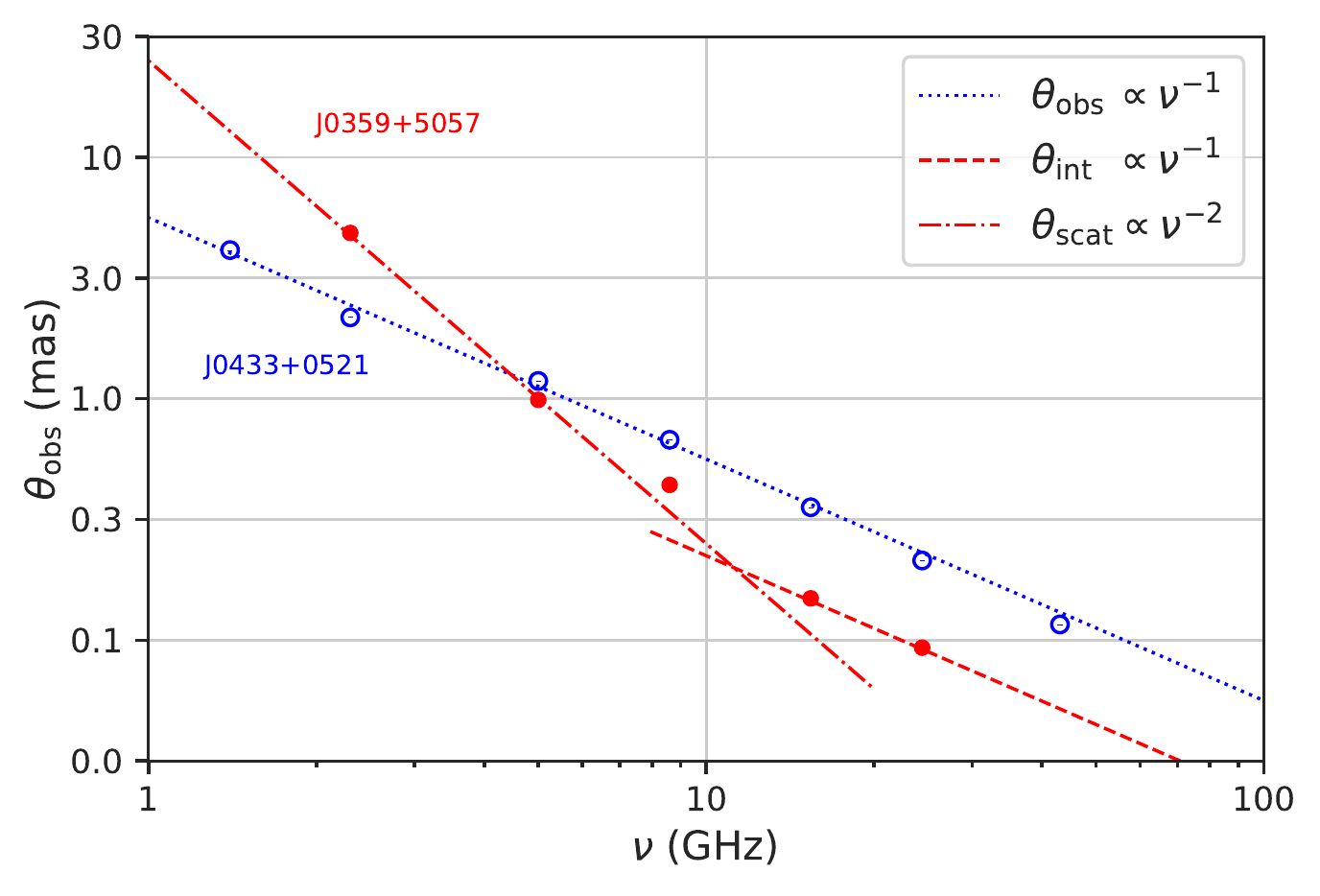}
    \caption{Example of the frequency dependence of the AGN core size for two sources: (1) the source with undetected scattering J0433+0521 (blue open circles) located at $(l,b)=(-169\fdg6,-27\fdg4)$ and (2) the scattered source J0359+5057 (red circles) located at $(l,b) = (150\fdg4,-1\fdg6)$. The size measurement error is not shown in the plot, since its value does not exceed the radius of the plot markers (red and blue circles). The dotted lines represent a fit according to the equations mentioned in the legend.}
    \label{fig:freq_vs_size}
\end{figure}

\autoref{fig:freq_vs_size} shows how the AGN core size changes depending on the observing frequency for sources used as two clearly different examples: J0433+0521, which is the source with undetected scattering, and J0359+5057, which is a highly scattered one. Therefore, as discussed and shown above, the observational data for J0433+0521 can be fitted by the size-frequency dependence $\theta_\mathrm{obs}\propto\nu^{-1}$. The frequency dependence for the scattered source J0359+5057 is best fitted by a sum of two spectral components with different slopes, $k_\mathrm{int}=1$ and $k_\mathrm{scat}=2$ in the logarithmic scale.

Using the multi-frequency observational data of the AGN core sizes in the Galactic plane ($|b|<10^\circ$) according to Equations~\ref{eq:formula1}--\ref{eq:formula1_2}, we can estimate the characteristic value of the scattering index $k_\mathrm{scat}$, which will correspond to the best fit of the intrinsic $\theta_\mathrm{int1}$ and scattered $\theta_\mathrm{scat1}$ sizes at 1~GHz setting $k_\mathrm{int}=1$. 
For each source, the average observed core size at a given frequency was taken.
We performed fitting for three frequency sub-groups.
The obtained values of the scattering index $k_\mathrm{scat}$ are summarised in \autoref{tab:table_k_scatter_results}. 
Note, in this section we used the distribution generated by bootstrap method to estimate the parameter confidence interval. Hence the errors of the values were estimated as the standard deviation of the obtained parameter distribution.

We compared the core size at 15~GHz from our two-component circular Gaussian modelfits with that from more detailed models from the 2-cm VLBA Survey \citep{2cmPaperIII} and its successor, the MOJAVE program \citep{MOJAVE_XVIII}, filtering out about 25~per~cent of cases when the core feature is fitted with delta-function and using geometric average in those cases (about 19~per~cent) where the core is an elliptical Gaussian. In total, we compared 5891 models of 503 sources and found that a distribution of the ratio $\theta_{\rm core}^{\rm 2comp}/\theta_{\rm core}^{\rm MOJAVE}$ peaks at 1 but has a heavier right tail that shifts the median to about 1.5. 
\begin{table*}
	\caption{Results of the correlation Kendall's $\tau$-test between values obtained in this work and the other measured characteristics of the Galaxy.}
	\centering
	\resizebox{\textwidth}{!}{
	\begin{tabular}{l|cccc|cccc|ccc} 
		\hline
		\hline
		 & & $|b| < 10^\circ$ &&& & $|b| > 10^\circ$ &&& & All sky &\\
		                      & $\tau$           & $N$ & $p$                 && $\tau$          & $N$    & $p$               && $\tau$          & $N$   & $p$ \\
		\hline
		
		$k$ -- $RM$                            & $0.122\pm0.031$  & 462 & $8.7\times10^{-5}$  && $0.065\pm0.013$ & 2614 & $5.9\times10^{-7}$  && $0.121\pm0.012$ & 3076  & $7.6\times10^{-24}$\\[2pt]
		
		$k$ -- $\theta_{\mathrm{NE2001}}$      & $0.206\pm0.030$  & 462 & $4.0\times10^{-11}$ && $0.071\pm0.013$ & 2614 & $4.4\times10^{-8}$  && $0.150\pm0.012$ & 3076  & $1.4\times10^{-35}$\\[2pt]
		
		$k$ -- $\mathrm{H\alpha}$              & $0.300\pm0.028$  & 462 & $6.2\times10^{-22}$ && $0.097\pm0.013$ & 2614 & $9.9\times10^{-14}$ && $0.173\pm0.012$ & 3076  & $3.4\times10^{-47}$\\[2pt]
		
		$\theta_{\mathrm{obs,2\,GHz}}$ -- $RM$ & $0.124\pm0.028$  & 510 & $2.8\times10^{-5}$ && $0.048\pm0.012$ & 3031 & $8.0\times10^{-5}$ && $0.111\pm0.012$ & 3541  & $4.8\times10^{-23}$\\
		
		$\theta_{\mathrm{obs,2\,GHz}}$ -- $\theta_{\mathrm{NE2001}}$ & $0.271\pm0.027$  & 510 & $6.3\times10^{-20}$ && $0.043\pm0.013$ & 3031 & $3.3\times10^{-4}$ && $0.140\pm0.012$ & 3541  & $5.3\times10^{-36}$\\
		
		$\theta_{\mathrm{obs,2\,GHz}}$ -- H$\alpha$& $0.369\pm0.028$  & 510 & $1.2\times10^{-35}$ && $0.084\pm0.013$ & 3031 & $3.3\times10^{-12}$ && $0.176\pm0.011$ & 3541  & $2.4\times10^{-55}$\\
		
		\hline
	\end{tabular}
	}
	\begin{tablenotes}
        \item Note: $k$ is the power-law index calculated by the two-frequency method (see \autoref{ch:two_freqs_method} for details) using the simultaneous observational data at 2~GHz and 8~GHz; $RM$ is the rotation measure; $\theta_\mathrm{NE2001}$ is the scattered AGN core size at 1~GHz obtained based the NE2001 model; $\mathrm{H\alpha}$ is the intensity of the $\mathrm{H\alpha}$ radiation in the Galaxy; $\theta_\mathrm{obs,2\,GHz}$ is the AGN core sizes measured at 2~GHz. $\tau$ is the Kendall's correlation coefficient, $N$ is the number of the sources, $p$ is the probability of chance correlation.
    \end{tablenotes}
	\label{tab:correlation_results}	
\end{table*}

This indicates that at relatively high frequencies where scattering is weak, we overestimate the core size to some degree applying the two-component approach to modelfit the observed source brightness distribution, which works better at lower frequencies. This leads to an underestimation of the $k$-index in the multi-frequency analysis. Thus, we conclude that the most robust and least biased estimate of $k$ in \autoref{tab:table_k_scatter_results} is for the 2-5-8~GHz fit which is based on the low-frequency data. For the same reason, we excluded the AGN core size measurements at 86~GHz from the multi-frequency analysis.

\begin{table}
	\caption{Scattering indices $k_\mathrm{scat}$ derived from different sets of the measured AGN core sizes at the Galactic plane.
	\label{tab:table_k_scatter_results}}
	\centering
	\begin{tabular}{lrc} 
		\hline
		\hline
		Subgroup & $N$ & $k_{\mathrm{scat}}$ \\
		(1) & (2) & (3)\\
		\hline
		
		All combinations            & 174  & $1.83\pm0.07$   \\[4pt]
		
		$\textbf{2,\,\,5,\,\,8}$ \textbf{GHz} & \textbf{130} & $\textbf{2.01$\pm$0.13}$   \\[4pt]
		
		$2,\,8,\,15$ GHz     & 58   & $1.83\pm0.07$   \\
		
		
		\hline
	\end{tabular}
	\begin{tablenotes}
       \item Columns are as follows: (1) frequency bands over which the subgroups of the sources were selected for fitting; (2) the number of the sources used to estimate the scattering index value; (3) the scattering index value.
       \item `All combinations' means that all  frequencies were used for each source, when measurements at more than three bands were available, but without 86~GHz. The most robust result is highlighted in bold (see \autoref{sec:section6_1}, paragraph 4 for details). 
    \end{tablenotes}
\end{table}

\begin{figure*}
    \centering
    \includegraphics[width=0.486\linewidth]{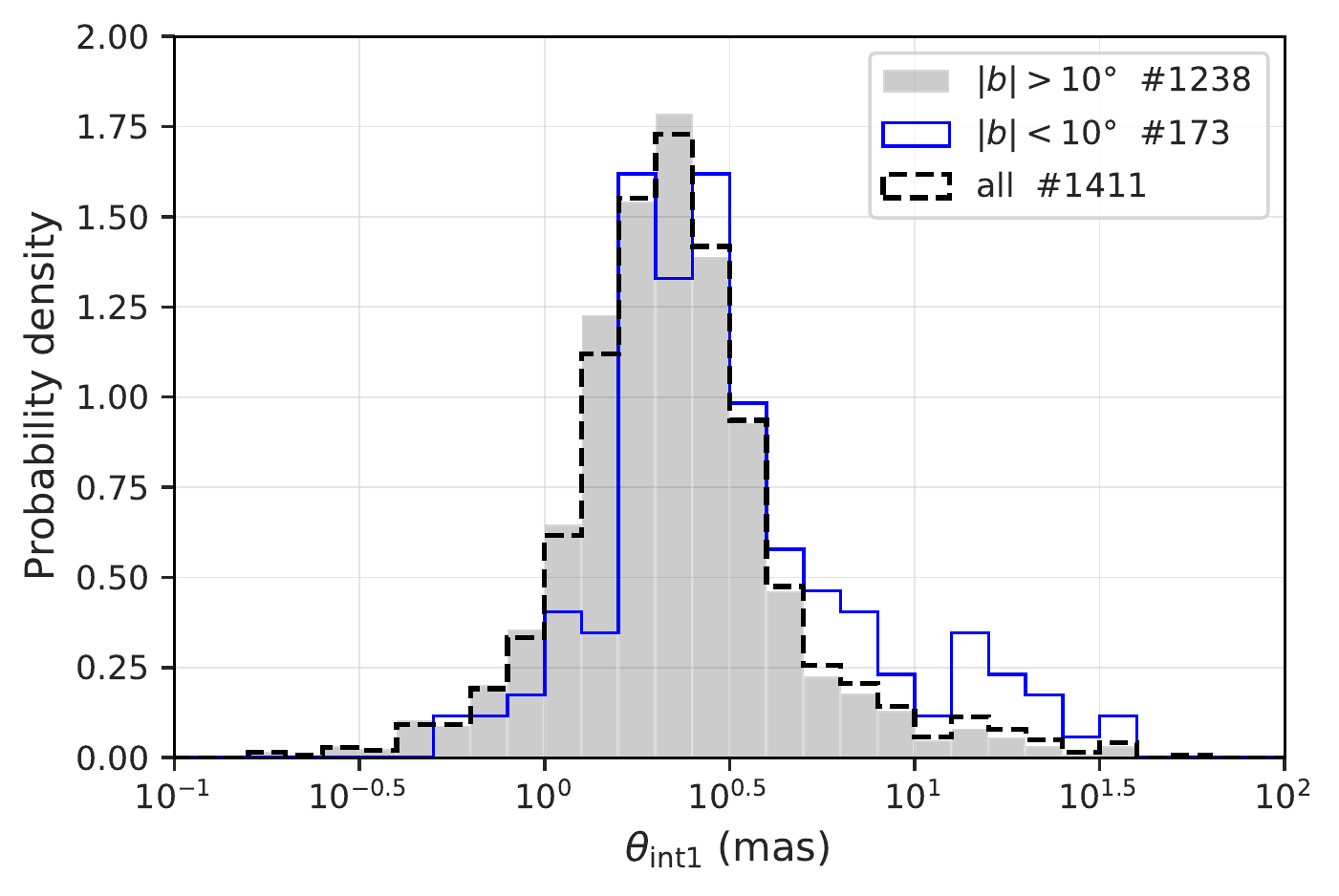}
    \includegraphics[width=0.474\linewidth]{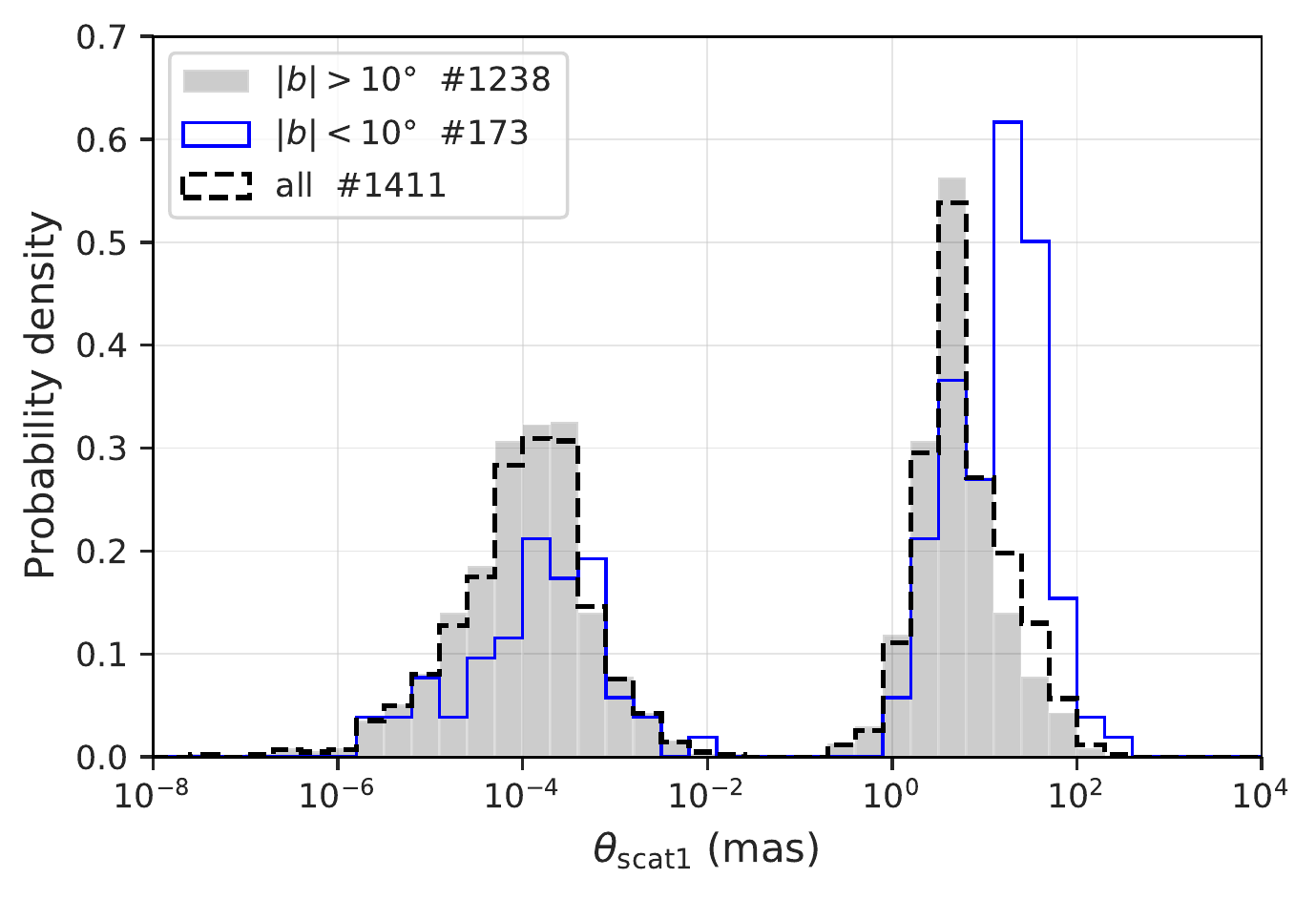}
    \caption{Histograms of the intrinsic ($\theta_\mathrm{int1}$, left) and scattered ($\theta_\mathrm{scat1}$, right) sizes of the AGN cores at 1~GHz obtained from the simulation. The blue contour histogram shows the size distribution for the sources in the Galactic plane. The grey solid histogram shows the size distribution outside the Galactic plane. The black dashed histogram shows the size distribution of the sources from the entire sky. The values given in the legend after the `\#' sign indicate the number of the sources for this set of the data.}
    \label{fig:theta_scat_and_int}
\end{figure*}

\begin{figure*}
    \centering
    \includegraphics[width=0.45\linewidth]{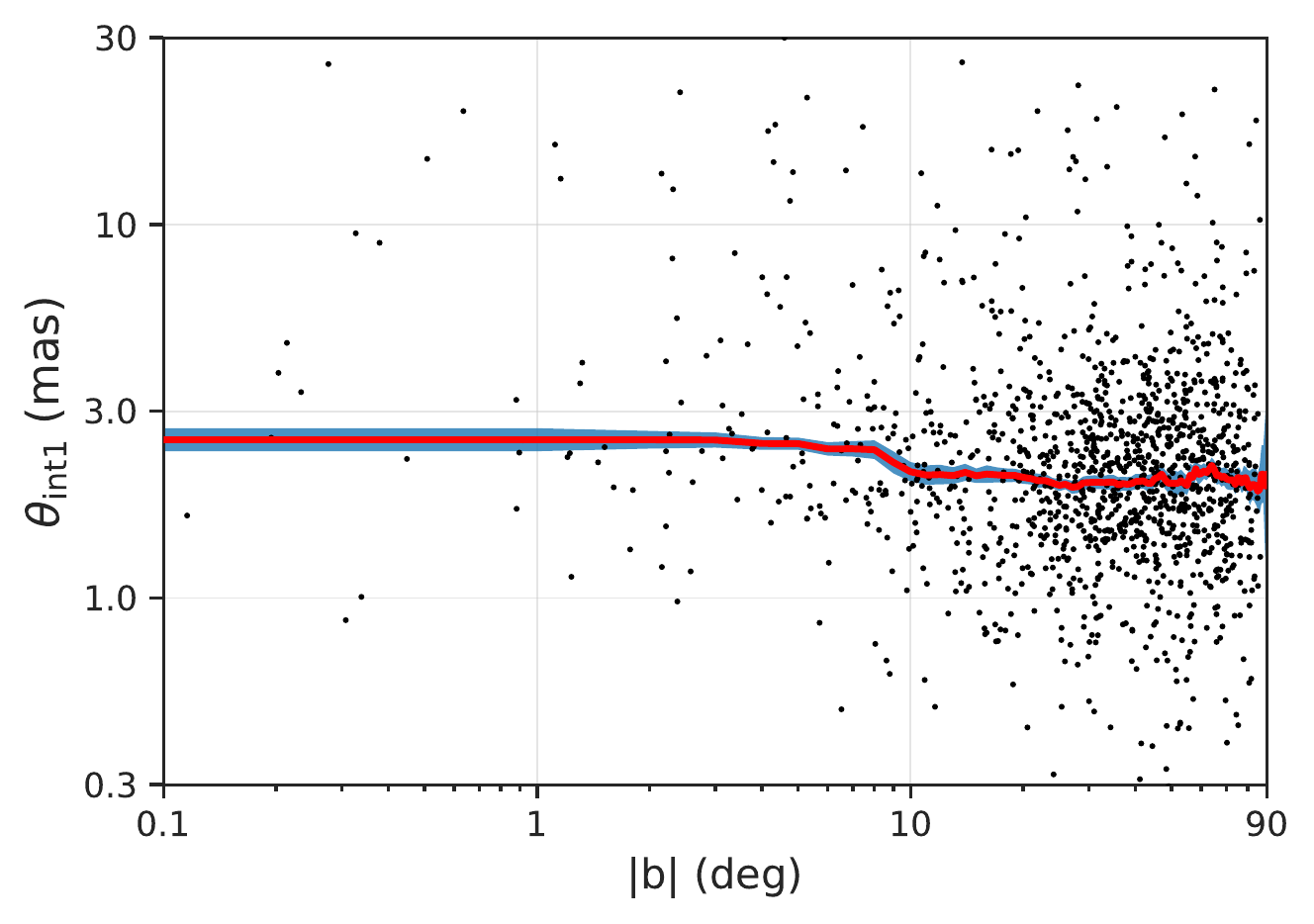}
    \hspace{0.5cm}
    \includegraphics[width=0.45\linewidth]{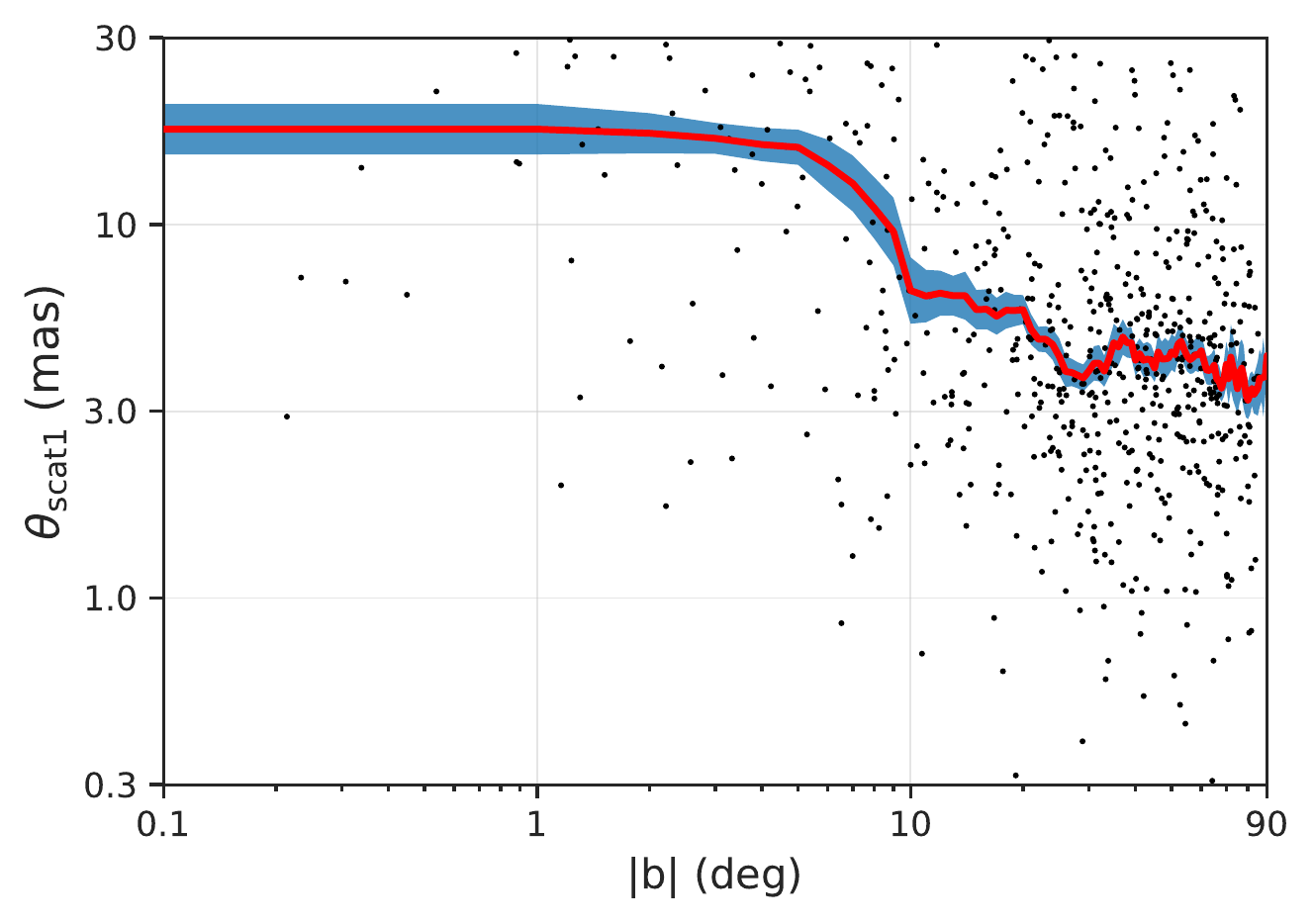}
    \caption{Intrinsic ($\theta_\mathrm{int1}$, left) and scattered ($\theta_\mathrm{scat1}$, right) AGN core sizes at 1~GHz obtained from the simulation, depending on the absolute value of the Galactic latitude. Each dot represents an individual source and is plotted only for AGN cores with the significant scattered component. The running median (the red curve) was taken over a range of $10^\circ$ of $|b|$. The blue shaded area shows the standard deviation of the median value of the angular size.}
    \label{fig:inn_scatt_vs_gallat}
\end{figure*}

Using the obtained values of $k_\mathrm{scat}$ from \autoref{tab:table_k_scatter_results}, we can separate the contribution of the intrinsic, $\theta_\mathrm{int1}$, and scattered, $\theta_\mathrm{scat1}$, component size to the observed one, $\theta_\mathrm{obs}$. To find $\theta_\mathrm{int1}$ and $\theta_\mathrm{scat1}$, we put the found value of the scattering index into Equations~\ref{eq:formula1}-\ref{eq:formula1_2}. For these calculations, the value $k_\mathrm{scat}=2.01\pm{0.13}$ was used due to the most robust combination of bands, as mentioned above. The sources from the entire sky that have more than three observing frequencies were selected for this analysis. The intrinsic, $\theta_\mathrm{int1}$, and scattered, $\theta_\mathrm{scat1}$, sizes were estimated at 1~GHz for 1411 AGN. The results are listed in \autoref{tab:table_thetas}. The obtained results are also presented in \autoref{fig:theta_scat_and_int}.

\begin{table}
	\caption{The intrinsic and scattered sizes of 1411 AGN VLBI cores at 1~GHz.}
	\centering
	\begin{tabular}{lrrrrr} 
		\hline
		\hline
		Name & $b$ & $\theta_{\mathrm{int1}}$& $\theta_\mathrm{int1}^\mathrm{err}$ & $\theta_{\mathrm{scat1}}$& $\theta_{\mathrm{scat1}}^\mathrm{err}$\\
		     & (deg) & (mas) & (mas)& (mas)& (mas) \\
		(1) & (2) & (3) & (4) & (5) & (6) \\
		\hline
        J0042$+$5708 & $-5.71$	& 0.86	  & 0.03    & 26.34     & 0.30 \\
        J0407$-$1211 & $-41.76$	& 2.23    & 0.02    & 3.19      & 0.11  \\
        J0805$+$6144 & 32.35	& 2.13	  & 0.02    & 1.91      & 0.24 \\
        J0809$+$3455 & 30.35	& 1.42	  & 0.04    & 5.25      & 1.27 \\
        J0909$+$4253 & 42.84	& 1.82	  & 0.03    & 1.06      & 0.47 \\
        J1217$+$3007 & 82.05	& 1.05	  & 0.05    & 3.64      & 0.15  \\
        J1305$-$1033 & 52.16	& 2.36	  & 0.02    & $<0.10$   & $\dots$ \\
        J1436$+$2321 & 65.95 	& 0.76    & 0.01    & $<0.10$   & $\dots$ \\
        J1728$+$3838 & 31.97	& 0.89    & 0.06    & 5.15      & 0.38 \\
        J2102$+$4702 &  0.34    & 1.01    & 0.07    & 14.19     & 0.49\\
		\hline
	\end{tabular}
    \begin{tablenotes}
        \item Columns are as follows: (1)~name of source; (2)~Galactic latitude; (3)~the value of calculated intrinsic size of the source; (4)~the intrinsic size calculation error estimated using the bootstrap method; (5)~the value of calculated scattered size of the source; (6)~the scattered size calculation error estimated using the bootstrap method.
        The full table is published in its entirety in the machine-readable format as supplementary material. Ten randomly selected records are shown here for guidance regarding its form and content.
    \end{tablenotes}
	\label{tab:table_thetas}	
\end{table}

The distribution of the intrinsic sizes (\autoref{fig:theta_scat_and_int}, left) has one peak. The median value of the intrinsic sizes in the Galactic plane (blue contour histogram) and outside of it (solid gray histogram) is 2.6 and 2.1~mas, respectively. There is a slight shift between the medians of the blue and gray histograms. We also analysed the dependence of the obtained intrinsic sizes on the absolute value of the Galactic latitude, which is shown in \autoref{fig:inn_scatt_vs_gallat}, left. As seen, the running median retains its value through the whole range of $|b|$ even within the Galactic plane. It is assumed that the intrinsic sizes do not depend on the Galactic latitude.

The distribution of the obtained scattered sizes of the AGN cores is presented in \autoref{fig:theta_scat_and_int}, right. For this distribution, a completely different picture is observed. The histogram has a two-peak structure.
The left peak is populated by the sources for which the scattered component of the measured size $\theta_\mathrm{scat1}$ is estimated to be very close to zero within the errors. This means that the contribution of scattering to the observed angular diameter is negligible; these are the sources with insignificant scattering. 
The right peak of the distribution corresponds to the sources for which the contribution of scattering is dominant or at least significant. For this peak, there is a significant shift between the distribution of the scattered sizes in the Galactic plane and outside of it. Compare blue and gray histograms in \autoref{fig:theta_scat_and_int}, right. The median values of scattered sizes in the Galactic plane (blue contour histogram) and outside of it (grey solid histogram) in \autoref{fig:theta_scat_and_int} are 16.1 and 4.8~mas, respectively. Therefore, we conclude that the contribution of the scattered component to the observed AGN core size in the Galactic plane is systematically larger than that of outside the Galactic plane, and this difference is significant. 
\autoref{fig:inn_scatt_vs_gallat} shows the dependence of the obtained scattered sizes on the absolute value of the Galactic latitude. We plotted only those sources for which the scattered component of the observed size is significant. The running median demonstrates a large increase in the median scattered sizes as it approaches and crosses the Galactic plane ($|b| = 10^\circ$).

\begin{figure*}
    \centering
	\includegraphics[width=\linewidth,trim=0cm 0.6cm 0cm 0cm]{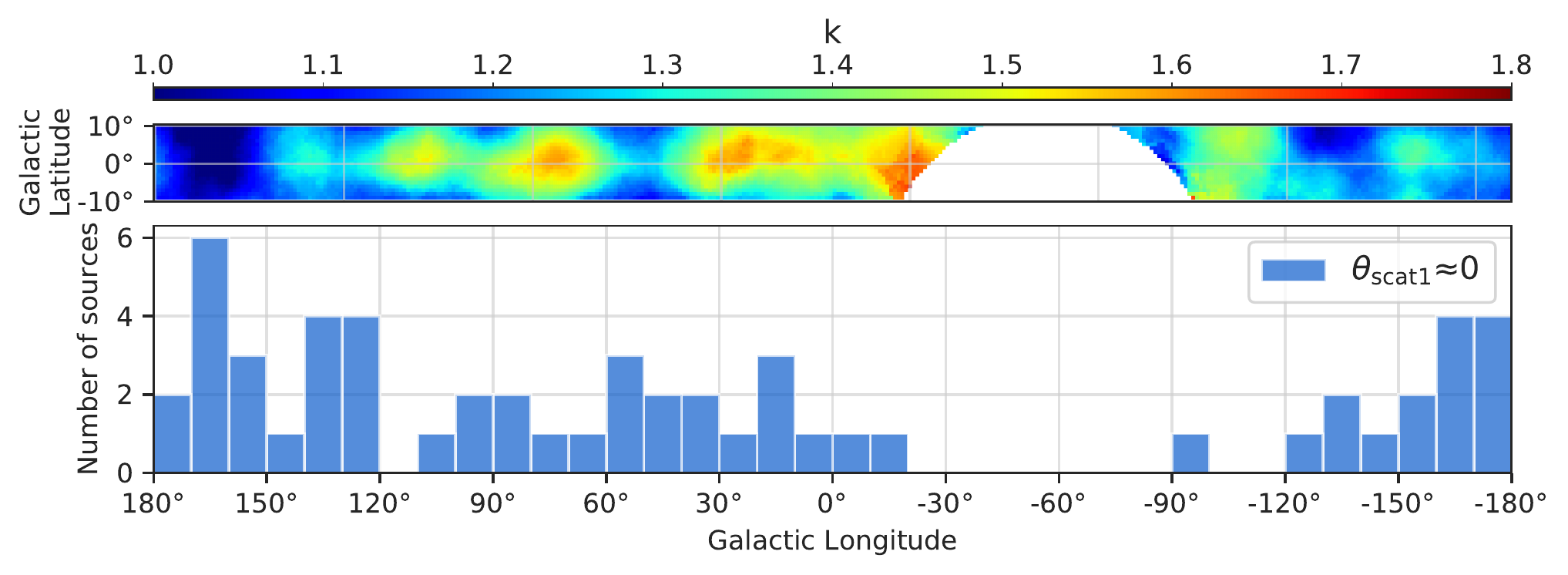}
    \caption{Histogram of the Galactic longitude for the sources with an insignificant scattered component of the observed AGN core size (bottom) against the map of the $k$ index distribution (top panel) corresponding to \autoref{fig:k28_map}, cropped at latitudes ($|b|=10^\circ$) and projected onto the plane. The axes of the Galactic longitudes in both panels coincide.}
    \label{fig:theta_zeros_vs_shpere}
\end{figure*}

We determined that for about 30~per~cent of AGN in the Galactic plane the contribution of the scattered component of the observed core size is negligible. It means that this fraction of the observed sources in the Galactic plane is not subject to significant scattering for frequencies down to at least 2 GHz. \autoref{fig:theta_zeros_vs_shpere} demonstrates that the sources with $\theta_{\mathrm{scat}1}\approx0$ are concentrated in the areas where the sources with insignificant scattering ($k\approx1$) are located. We note that the results presented on two panels in \autoref{fig:theta_zeros_vs_shpere} being obtained using different methods and data sets, are in good agreement with each other. Therefore, we can conclude that most of the sources with insignificant scattering are located in the region of the Galactic anti-centre ($|l|>150^\circ$).

\begin{figure}
    \centering
	\includegraphics[width=0.95\linewidth]{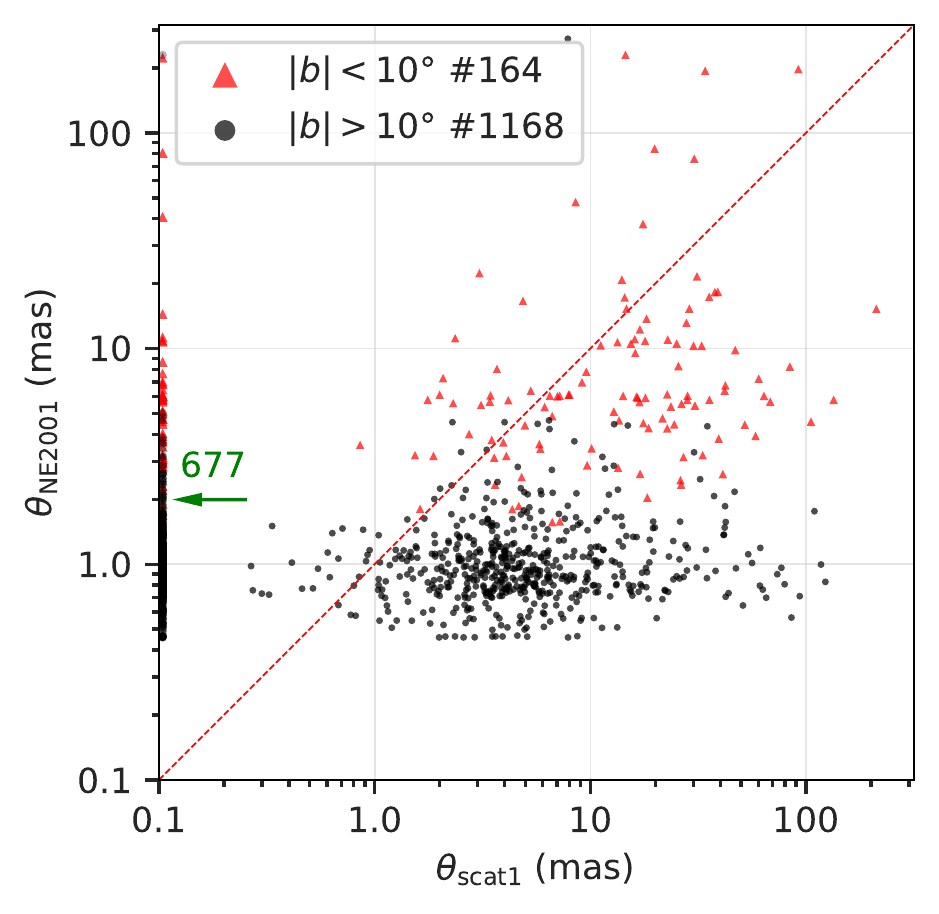}
    \caption{Comparison of the results obtained from our modelling of the scattered AGN core sizes $\theta_\mathrm{scat1}$ with the sizes predicted by the NE2001 model  at 1~GHz. See \autoref{ch:corellation_tests} for details. The green arrow points to 677 missing sources lying outside the plot (their y-axis positions are marked). The values given in the legend after the `\#' sign indicate the number of the sources for this dataset.}
    \label{fig:NE2001_vs_theta_scatt}
\end{figure}

The free electron density distribution in the Galaxy model NE2001 \citep{NE2001} is mainly based on radio observations of pulsars which provide the information about the properties of the local ionized interstellar medium. Pulsars being Galactic radio sources are not optimal for studying large-scale scattering properties. The pulsar data can provide reliable information only in the Galactic plane. Nevertheless, we performed a comparison of the estimated scattered sizes $\theta_\mathrm{scat1}$ obtained in our work, with the scattered sizes obtained on the basis of the NE2001 model ($\theta_\mathrm{NE2001}$) at 1~GHz. In \autoref{fig:NE2001_vs_theta_scatt} we show the sources lying in the Galactic plane and outside of it in different colors. The scattering sizes determined on the basis of the NE2001 model in the region $|b|>10^\circ$ are mainly concentrated around $\theta_\mathrm{NE2001} \approx 1$~mas (black circles), but according to our results the $\theta_\mathrm{scat1}$ sizes of these sources cover the range from 0.1~mas to more than 10~mas. The extragalactic objects, the path to which lies through the Galactic plane (red triangles), have a better agreement between the compared sizes. The red dots are scattered around the line of equality $\theta_\mathrm{scat1}=\theta_\mathrm{NE2001}$. Nearly half of the sources (677) lie outside the plot, we mark them along the y-axis. These are the sources for which the scattering does not dominate according to our calculations, $\theta_\mathrm{scat1}\approx0$, but the NE2001-model sizes are distributed from 0.1~mas to about 8~mas. Thus we conclude that the scattering of these sources predicted by the NE2001 model was not confirmed by the results of our estimates made on the basis of multi-frequency VLBI AGN data.

\subsection{Modeling observed core size distributions at 2~GHz and 8~GHz}

In this section we describe an alternative method for estimating the scattering index $k_\mathrm{scat}$ using the AGN VLBI core size measurements at two frequencies. For simplicity, we here assume that sources seen outside the Galactic plane ($|b|>10^\circ$) are not subject to scattering at all. In \autoref{fig:k_28_hists}, we showed that the distribution of the observed AGN core sizes in the Galactic plane ($|b|<10^\circ$) contains both sources with insignificant (about 30~per~cent) and significant scattering. Therefore, the size distribution of the sources from this area contains the contribution of both types of sources.

We also assume that the observed size distribution of the scattered sources could be fitted with a generalization of half-Gaussian distribution or half-Student's distribution. We performed an iteration over all parameters of these distributions. For example, for the Student's function: $do\!f$ and $\sigma$ are degrees of freedom and width of the distribution, respectively. We generated a random sample of sizes, and then added in quadrature to the observed sizes of the sources outside the Galactic plane (the sources with insignificant scattering), that is $\theta_{\mathrm{int}}$ according to \autoref{eq:formula1}. The result of these steps is the distribution $\theta_\mathrm{fit}$, which we will compare with the distribution obtained from the observational data for the sources in the Galactic plane $\theta_\mathrm{obs}$ ($|b|<10^\circ$). If these distributions are similar, the selected analytical distribution with its parameters is an appropriate description of $\theta_\mathrm{scat}$ distribution at a given frequency.

\begin{figure}
    \centering
    \includegraphics[width=\linewidth]{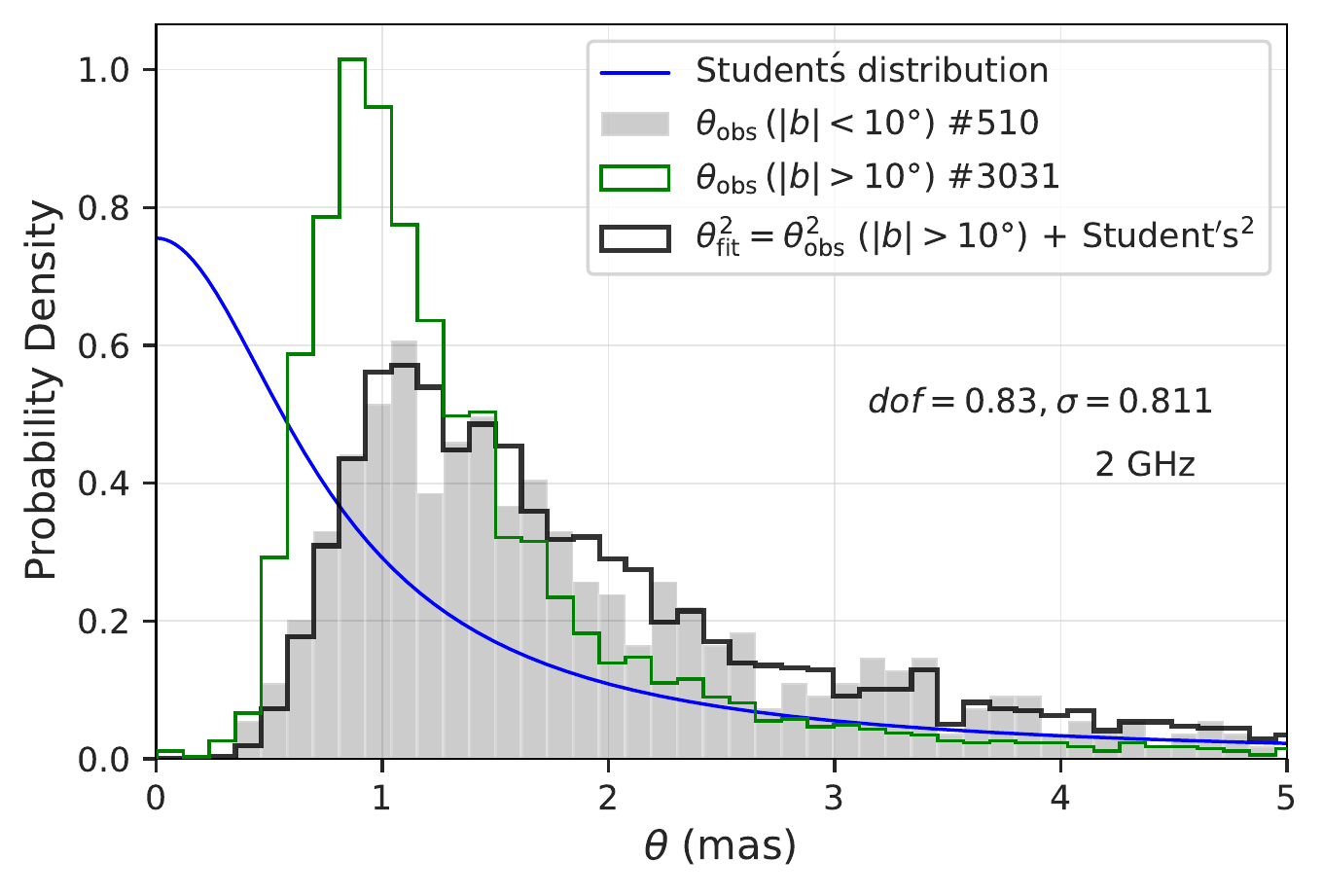}
    \includegraphics[width=\linewidth]{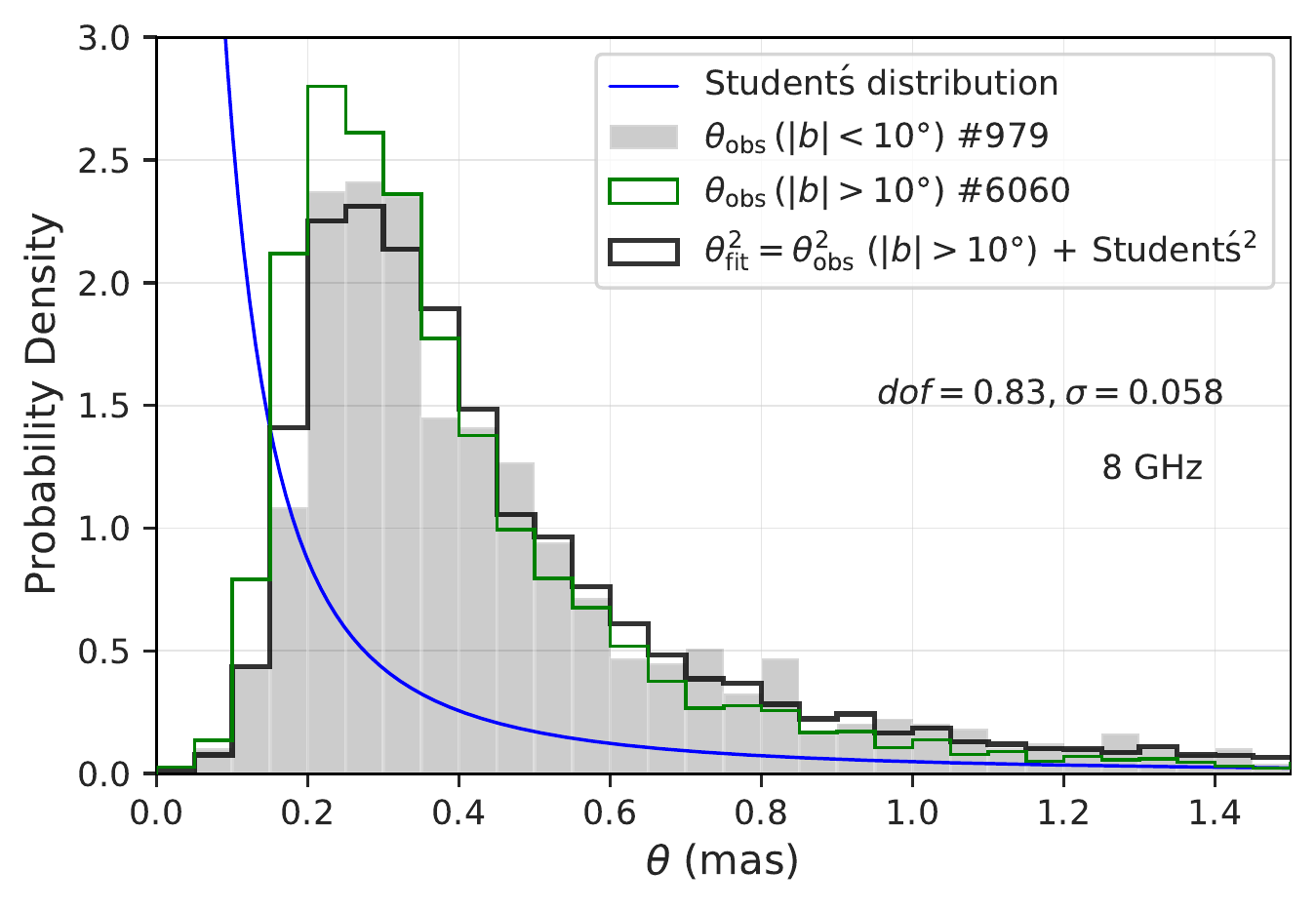}
    \caption{
    Histograms of the observed AGN core sizes $\theta_\mathrm{obs}$. The gray histogram represents the AGNs in the Galactic plane while the green contour histogram shows the distribution outside the plane. The blue curve is the distribution of the scattered sizes at a given frequency $\theta_{\mathrm{scat}}$ (see \autoref{eq:formula1}) applying half-Student's distribution. The fitting parameters for half-Student's function are shown on the right side of the plots; $do\!f$ is degrees of freedom, $\sigma$ is the standard deviation. The black contour histogram depicts total distribution of the source sizes $\theta_\mathrm{fit}$, which is the squared sum of half-Student's distribution and the distribution of the sources with insignificant scattering ($|b|>10^\circ$). The values given in the legend after the `\#' sign indicate the number of the sources for a given dataset. 
    \label{fig:size_distribution_fitting}
    }
\end{figure}

We minimize the Kolmogorov-Smirnov (KS) test statistic between the observed and fitted size distributions.
The smallest values of these statistics, i.e., the greatest similarity of distributions, is obtained when we used the Student's distribution to fit the scattered size $\theta_{\mathrm{scat}}$. \autoref{fig:size_distribution_fitting} shows the data fitting results at two frequencies, 2~GHz and~8 GHz (black contour histograms).

We searched for the best widths $\sigma$ at 2~GHz and 8~GHz with a fixed value of the degree of freedom, $do\!f$, for both frequencies. Thus, the found parameters of $\sigma$ for 2~GHz and 8~GHz can be used to derive the value of the scattering index, similarly to the two-frequency method (see \autoref{ch:two_freqs_method} for details). We reached the maximum similarity with the following parameters of the Student's function: $do\!f = 0.83\pm0.06$, $\sigma= 0.811\pm0.077$ for 2~GHz, $\sigma= 0.058\pm0.008$ for 8~GHz. Thus, using this alternative method, the scattering index $k_\mathrm{scat} = 2.00 \pm 0.05$ was inferred. All the errors were estimated as standard deviations of the obtained by bootstrap parameter distributions.

\subsection{The special case of Sagittarius~A$^\ast$}

\cite{2018ApJ...865..104J} have shown that for Sagittarius~A$^\ast$, the most heavily scattered source on the sky, the power-law index for density fluctuations $\beta<3.47$, which is shallower than expected for a Kolmogorov spectrum ($\beta=11/3$). At the same time they found $\lambda^2$ dependence of the angular size ($k=2.0$) at $\lambda>1$~cm, as the scattering is in a strong regime, with the diffractive scale smaller than the dissipation scale of the turbulence. In this case, the source angular size follows the $\lambda^2$ scaling regardless of a spatial spectrum slope. In our case, the sources are more weakly scattered, and the obtained value of $k$ indicates that the spatial spectrum of inhomogeneities is steeper ($\beta\simeq 4$) than that for a Kolmogorov turbulence.

\section{Summary}
\label{s:summary}

We used the largest to~date number of experimental VLBI data from multi-frequency observations of AGNs to study the ISM scattering properties in the Galaxy. We analyzed the dependencies of the observed AGN core sizes on the Galactic latitude and found significant angular broadening of the measured core sizes for the sources seen through the Galactic plane. This effect is especially strong at low frequencies, particularly at 2~GHz. We established that scattering screens containing density fluctuations of hot plasma are concentrated mainly in the Galactic plane. Outside the Galactic plane, we did not detect regions with strong scattering. We created the sky distribution maps of the measured AGN core sizes at 2, 5, and 8~GHz, marking the distribution of scattering screens in the Galaxy.

We calculated the power-law index $k$ from the frequency dependence of the AGN core size $\theta\propto\nu^{-k}$ derived from simultaneous observations at 2~GHz and 8~GHz for 2614 sources. The mean value $k = 1.02\pm0.01$ derived for the sources outside the Galactic plane is in good agreement with the theoretical prediction $\theta\propto\nu^{-1}$ for the AGN cores with undetected scattering with synchrotron self-absorption. For the scattered sources, we obtained a characteristic value $k = 1.60\pm0.02$ ignoring the contribution of the intrinsic structure, hence it is less than the predicted $k \simeq 2$. Using the $k$-index values calculated between 2~GHz and 8~GHz, we constructed the first detailed sky distribution map of scattering properties in the Galaxy based on VLBI experimental data. 

The regions of the Galaxy characterized by a high $\mathrm{H\alpha}$ radiation intensity show a significant spatial correlation with the areas of strong scattering. One of them is positionally associated with the Cygnus constellation region, which contains active star-forming regions and the supernova remnant W78. In the locations of Taurus~A, Vela and Cassiopeia~A supernova remnants, as well as in the location of Orion nebula (M42), we also found an increase of the scattering strength of the ISM. The region with the strongest scattering is the Galactic centre, which extends in the Galactic plane at $-20^\circ\leqslant l\leqslant20^\circ$.

Using the AGN VLBI core sizes derived from multi-frequency data, we separated the contribution of the intrinsic and scattered sizes to the observed angular diameter for 1411 AGN. As expected, the contribution of scattered components of the observed size for the sources in the Galactic plane is systematically larger than for those observed outside the Galactic plane. We found that about 30~per~cent of the AGN observed in the Galactic plane are not subject to scattering. This reflects that the interstellar medium of our Galaxy is highly inhomogeneous. Most of the sources with insignificant scattering are located in the direction of the Galactic anti-centre.

Applying different methods to derive the power-law scattering index, we obtained virtually the same values $k_\mathrm{scat} \simeq 2.0$. This strongly supports the Gaussian screen model. At the same time, we do not exclude that new targeted observing campaigns to study main scattering screens individually may deliver for some screens different results that will be more consistent with the Kolmogorov turbulence model.

\section*{Acknowledgements}

We thank the anonymous referee and Eduardo Ros for comments which have helped to improve the manuscript as well as Elena Bazanova for language editing. 
We thank Michael Johnson for providing the data of the scattered size estimated based on the NE2001 model. 
We are grateful to the teams referred to in \autoref{ch:size_measurements} for making their fully calibrated VLBI FITS data publicly available and Leonid Petrov for maintaining the database with these data. 
This research was supported by the Russian Science Foundation project 21-12-00241. 
This research has used the MOJAVE database maintained by the MOJAVE team \citep{2018ApJS..234...12L}. This study has also used the VLBA data from the Blazar Monitoring Programs BEAM-ME and VLBA-BU-BLAZAR, funded by NASA through the \emph{Fermi} Guest Investigator Program. 
This research has made use of NASA’s Astrophysics Data System.

\section*{Data Availability}

The analysis is based on the data compiled in the Astrogeo database\footnote{\url{ http://astrogeo.org/vlbi_images/}} that collects VLBI observations. The dataset contains 17\,474 sources observed from 1994 to 2021, more than $100\,000$ individual observations. The majority of them were performed at 2, 5, 8 or 15~GHz. Details are presented in \autoref{ch:size_measurements}. The data underlying this study is available in the paper and in its online supplementary materials. 


\bibliographystyle{mnras}
\bibliography{article}


\bsp 
\label{lastpage}
\end{document}